\documentclass[lettersize,journal]{IEEEtran}
\usepackage{amsmath,amsfonts}
\usepackage[ruled,vlined]{algorithm2e}
\usepackage{array}
\usepackage[caption=false,font=scriptsize,labelfont=sf,textfont=sf]{subfig}
\usepackage{textcomp}
\usepackage{stfloats}
\usepackage{url}
\usepackage{verbatim}
\usepackage{graphicx}
\usepackage{cite}
\usepackage{paralist}

\def\BibTeX{{\rm B\kern-.05em{\sc i\kern-.025em b}\kern-.08em
    T\kern-.1667em\lower.7ex\hbox{E}\kern-.125emX}}

\usepackage{balance} %

\usepackage{booktabs}
\usepackage{multirow}

\usepackage{makecell}

\usepackage{enumitem}

\usepackage{tikz}
\usetikzlibrary{math,fit,arrows,arrows.meta,shapes,shapes.geometric,positioning,calc,shadows,trees,decorations.pathreplacing,calligraphy}

\usepackage[hidelinks]{hyperref}
\usepackage{cleveref}

\usepackage{adjustbox}

\newcommand{\mtx}[1]{\mathbf{#1}}
\newcommand{\vect}[1]{\boldsymbol{\mathbf{#1}}}
\newcounter{rqno}

\newcommand{\code}{\texttt}

\newcommand{\metric}{\texttt}

\providecommand{\shortcite}[1]{\cite{#1}}
\providecommand{\toprule}[1]{\hline{#1}}
\providecommand{\midrule}[1]{\hline{#1}}
\providecommand{\bottomrule}[1]{\hline{#1}}

\usepackage{xspace}
\def \numprojs {{10}\xspace}
\def \numapacheprojs {{6}\xspace}
\def \mplus  {(\texttt{+})}
\def \mminus {(\texttt{-})}
\def \meq    {(\texttt{=})}

\begin{document}
\title{Deep Incremental Learning of Imbalanced Data for Just-In-Time Software Defect Prediction}

\author{Yunhua Zhao, Hui Chen
	\thanks{Y. Zhao is with Department of Computer Science, CUNY Graduate Center,
	365 5th Avenue, New York, NY, 10016. (e-mail: yzhao5@gradcenter.cuny.edu)}
	\thanks{H. Chen is with Department of Computer and Information Science,
	CUNY Brooklyn College, 2900 Bedford Ave, Brooklyn, NY 11210 and with
	Department of Computer Science, CUNY Graduate Center,
	365 5th Avenue, New York, NY, 10016. (e-mail: huichen@ieee.org)}
}
\maketitle

\begin{abstract} 
  This work stems from three observations on prior Just-In-Time Software Defect
Prediction (JIT-SDP) models. First, prior studies treat the JIT-SDP problem
solely as a classification problem.  Second, prior JIT-SDP studies do not
consider that class balancing processing may change the underlying
characteristics of software changeset data. Third, only a single source of
concept drift, the class imbalance evolution is addressed in prior JIT-SDP 
incremental learning models.

We propose an incremental learning framework called CPI-JIT for JIT-SDP.
First, in addition to a classification modeling component, the framework
includes a time-series forecast modeling component in order to learn temporal
interdependent relationship in the changesets.  Second, the framework
features a purposefully designed over-sampling balancing technique based on
SMOTE and Principal Curves called SMOTE-PC. SMOTE-PC preserves the underlying
distribution of software changeset data.

In this framework, we propose an incremental deep neural network model called
DeepICP.  Via an evaluation using \numprojs software projects,  we show that:
1) SMOTE-PC improves the model's predictive performance; 2) to some software
projects it can be beneficial for defect prediction to harness temporal
interdependent relationship of software changesets; and 3) principal curves
summarize the underlying distribution of changeset data and reveals a new
source of concept drift that the DeepICP model is proposed to adapt to.

\end{abstract}


\section{Introduction}
Change-level or more often Just-In-Time Software Defect Prediction (JIT-SDP) is
to assess the risk of defects, colloquially bugs introduced by code changes in
the software~\cite{kamei2012large, fukushima2014empirical}.
Prior JIT-SDP models are often formulated as classification problem where we
give a ``{\em clue}'' to the trained model, the clue is the features about the
target changeset for which we make the prediction. Our work are motivated from
two primary observations on prior JIT-SDP models.

First, the ``{\em clue}'' are a representation of a changeset whose
defect-inducing status we are seeking. However, the clue is given in isolation,
completely removed from its historical context, i.e., the software changesets
preceding to the target changeset. Our reasoning is as follows.
\begin{inparaenum}
	\item The changesets are not independent events. During software development
	process, developers make and commit code changes one after another,
	resulting in sequences of code changes where some new code changes depend
	on old changes.  For instance, change couplings and genealogies exist among
	software changes~\cite{d2009relationship,herzig2013predicting}.  The
	dependent relationship of changesets is a clear contrast to the implied
	assumption of batch learning JIT-SDP models that in effect treat software
	changesets as independent events for the fact that there are no assumed
	order among software changesets and the changesets are often shuffled during
	training.
	\item Furthermore, the changesets over time represent the evolutionary
		history of software that we should take advantage of in JIT-SDP.
		This is because the evolutionary history has already been shown to be
		useful to predict a software system's future evolution, as such, can
		sometimes successfully serve as a starting point to determine what
		future engineering activities may be~\cite{d2009relationship}.  A
		category of such works is to predict change-proneness of software
		modules (e.g., classes in a software system)~\cite{elish2013suite};
		another is to foretell future refactoring
		activities~\cite{ratzinger2007mining}, and yet another is to determine
		future requirement changes~\cite{shi2013learning}.  As indicated in
		prior studies, changes to some software modules are more defect prone,
		refactoring changes and defects have often an inverse correlation, and
		some requirement changes are more complex~\cite{ratzinger2008relation}.
		Given these, the past changes likely have a bearing on what changes may
		occur in the future and their risk of introducing defects.
\end{inparaenum}

Second, prior works rarely consider that the preprocessing steps that balances
the training data sets alter underlying distribution of the data, which can
pose a threat to predictive performance of JIT-SDP models. Our work leads to
a proposal of balancing technique that preserves underlying data distribution,
which also avails us an opportunity to deal with types of concept drifts other
than the evolution of class imbalance status.

Software changesets are imbalanced with regard to their defectiveness
status, i.e., for some software projects, defect-inducing changesets are far
less than clean changesets.
Two types of
balancing techniques are often seen in prior JIT-SDP studies, and they are,
over-sampling approaches and under-sampling approaches~\cite{kamei2012large,
tan2015online, pornprasit2021jitline}.  The former like the Synthetic Minority
Over-Sampling technique (SMOTE) add artificially generated software changesets
of the minority class to the training data set, and the later remove software
changesets from the majority class of the training data set.  A question that
follows is, do these class imbalancing processing techniques alter the software
system's evolutionary history? Another question is, do they change the
underlying characteristics or the distribution of the changeset data?  JIT-SDP
models work because they learn the regularities in the historical changeset
data and their defectiveness status.
It begs the question, does the change of
the characteristics due to the class balancing processing also alter in the
training data set the very regularities with which the models predict
defectiveness status of software changes, and if so, does it impact negatively
the model's predictive performance on the test data set?

Recent
studies~\cite{tan2015online, cabral2019class, tabassum2020investigation}
suggest that incremental learning of which online learning is a special case
should be a more realistic learning setting to build JIT-SDP models.  The
incremental learning setting allows a JIT-SDP model to learn to predict
defective changes from new training changeset instances and their defectiveness
status gradually as they arrive, renew and adapt the model over time, which
offset the obstacles discussed in the above.  Although prior studies have
examined how the online models address one source of concept drift, the
evolution of the class imbalance status of software changeset
data~\cite{cabral2019class}, we wonder, is there any other form of concept
drifts that we should design an incremental JIT-SDP model to adapt to? Our
work provides an evidence to an affirmative answer to these questions.

Motivated by these observations, we propose an incremental learning
framework for JIT-SDP. \begin{inparaenum}
\item 
	We develop a new balancing technique called SMOTE-PC to preserve the
	characteristics of software change data while balancing the data.
\item 
	The framework encompasses three components, SMOTE-PC, a time-series forecast
	modeling component, and a classification modeling component.  The forecast
	component learns feature representations from the temporal interdependent
	relationship of the software changesets that precede the target changeset
	and these preceding changesets constitutes a context of the target changeset.
\end{inparaenum}

We assess and evaluate our hypotheses using \numprojs software projects. 
Our results confirm that:
\begin{inparaenum}
\item	SMOTE-PC does not alter underlying data distribution as much as SMOTE
	does and can help boost JIT-SDP models' predictive performance;
\item historical changesets exhibit temporal interdependent relationship;
\item although noisy, the information contained in the changesets preceding to
	the target changeset can benefit some software projects for defect
	prediction.
\end{inparaenum}
Following these, in the proposed framework, we design an incremental JIT-SDP
model called DeepICP taking advantage of advances in deep neural networks. We
show that for some software projects and on average the model can learn
incrementally and update as new changes arrive to adapt to concept drifts
broadly defined as the changes in the underlying data distribution of the
changeset data. 

The rest of this paper has the following organization. In Section~\ref{sec:rqs}, we
present a list of research questions and a summary of the research
contribution. To provide a context and a background for this study, we discuss
related works including prior studies in JIT-SDP, incremental and online
learning in JIT-SDP, and  principal curves and class imbalance processing in
Section~\ref{sec:related}. Section~\ref{sec:data} is a description of the
evaluation data set of \numprojs software projects and data preparation steps.
Sections~\ref{sec:relation} to~\ref{sec:drift} are to examine empirically our
hypotheses. In particular, Section~\ref{sec:relation} shows that software
changesets have temporal inter-dependent relationship and the temporal
relationship contains useful information to support defect prediction of
software changes, and Section~\ref{sec:drift} is an exploration of SMOTE and a
representation of software changeset concept drift using principal curves,
which lays a foundation for our proposal of SMOTE-PC.
Section~\ref{sec:smotepc} details our proposal of SMOTE-PC and our examination
of this class imbalance processing technique. The discussions in these sections
lead to the proposal of the new incremental learning framework for JIT-SDP,
which we discuss and evaluate in Section~\ref{sec:deepicp}. Finally, we discuss
the threats to validity in Section~\ref{sec:threats}  and conclude this article
in Section~\ref{sec:sum}.

\section{Research Questions and Contributions}
\label{sec:rqs}

In this work, we are to answer the following research questions. 

\begin{enumerate}%
	\item (RQ.1) Do software changesets exhibit temporal interdependent
		relationship that we can take advantage to design an incremental JIT-SDP
		model?

	\item (RQ.2) Do software changesets have concept drift beyond class imbalance
		evolution to which we should design incremental JIT-SDP models to adapt to?

	\item (RQ.3) Can SMOTE-PC, a purposefully designed class balancing processing
		technique preserve the underlying distribution of software
		changeset data to the benefit of defect prediction?

	\item (RQ.4) How do SMOTE-PC and concept-drift impact the DeepICP, a JIT-SDP
		model of simple realization of the proposed incremental learning framework
		to predict defective changes? 

\end{enumerate}

By answering the above research questions, we make the following contributions
to the study of JIT-SDP.
\begin{enumerate}
	\item By examining \numprojs software projects, we provide an empirical
		evidence to confirm the existence of temporal interdependent relationship
		among software changesets.  Due to this, we hypothesize
		the changesets
		preceding to the target changeset contain
		useful information for us to foretell characteristics of the target changeset
		without even seeing it.  In order to capture such information to benefit
		defect prediction, in contrast to prior JIT-SDP models, when we predict whether
		a target software changeset is defective, the clue about the target changeset
		given to the JIT-SDP models should include both the features of the target
		changesets and those of its preceding changesets where the preceding
		changesets provide a context of the target changeset. Our results show
		that in this way it is possible to harness the
		information from the relationship to help predict defectiveness status of
		software changes for some software projects.

	\item 
		Prior JIT-SDP studies have indicated that we should consider the
		longitudinal variations of the characteristics of software changeset data
		or concept drift when designing JIT-SDP models.  Prior studies have
		proposed online learning JIT-SDP models that adapt to one source of concept
		drift, the class imbalance evolution~\cite{cabral2019class,
		tabassum2020investigation}.  We propose to characterize the concept drift
		using the underlying distribution of software changeset data. Since
		principal curves are a nonlinear summary of the distribution of
		data~\cite{hastie1989principal, kegl2000learning}, it becomes a useful tool
		to recognize concept drifts beyond the class imbalance evolution.

	\item 
		JIT-SDP is a class imbalance problem~\cite{kamei2012large, tan2015online,
		cabral2019class}.  Most prior JIT-SDP studies approach the class imbalance
		problem by a routine class imbalance processing~\cite{kamei2012large}.
		Recent incremental and online JIT-SDP models begin to treat the class
		imbalance evolution as a source of concept drift and adapt the models to
		the concept drift~\cite{cabral2019class, tabassum2020investigation}.
		However, none of the prior studies consider that the class imbalance
		processing may alter the very regularities with which the JIT-SDP models
		predict defectiveness status of software changes.
		We propose a new class balancing technique called SMOTE-PC that
		extends the Synthetic Minority Over-Sampling technique (SMOTE) using
		Principal Curves.  Principal curves are non-linear curves that summarize the
		data distribution~\cite{hastie1989principal, kegl2000learning}.  Using
		principal curves, we characterize underlying distribution of software
		changeset data and design SMOTE-PC to  preserve the underlying data
		distribution.  This is motivated from our belief that in order to  improve a
		JIT-SDP model's predictive performance, we should preserve the underlying
		characteristics of software change data while balancing the data because the
		regularities in the data with which a JIT-SDP model predicts defectiveness
		status of software changes would otherwise also change.
		Our experiments indicate that a
		JIT-SDP model with SMOTE-PC can have better predictive performance than
		that with ordinary SMOTE in terms of F1 score.

	\item Finally, we propose an incremental learning framework for JIT-SDP.  We
		argue that in the framework it can be beneficial to include
		concept-preserving class imbalancing processing technique like SMOTE-PC, to
		include a forecast model component to extract from preceding changesets the
		features about the target software changeset, and to combine these with a
		classification component. We provide a JIT-SDP model called DeepICP, a
		model of simple realization in the framework. With this model, we examine
		its defect prediction ability and its ability to adapt to concept drift in
		the software changeset data.

\end{enumerate}

\section{Background and Related Work}
\label{sec:related}
Relevant to the proposed work, prior research has examined change-level
software defect prediction and in this context studied a broad set of models
and problems.  This section aims to situate the proposed work by discussing
most relevant prior JIT-SDP studies including the origin of JIT-SDP studies,
the evolution of modeling approaches, the incremental and online learning and
the concept drift problem in JIT-SDP. Additionally, we also introduce the
concept of principal curves. 

\subsection{JIT-SDP}
Mockus and Weiss~\shortcite{mockus2000predicting}, Aversano et
al.~\shortcite{aversano2007learning}, and Kim et
al.~\shortcite{kim2008classifying} are the early studies that introduce the
concept of changel-level Software Defect Prediction (SDP). Kamei et
al.~\shortcite{kamei2012large} carry out a large scale study on change-leve
SDP. The work establishes the benefits of this type of SDP for software quality
assurance, distills prior software quality studies into 14 software change
metrics, proposes two lines of defect prediction models, i.e., defectiveness
status prediction and effort-aware prediction, and examines the predictive
performance and the interpretations of the models using the software change
metrics of  a large set of open-source and proprietary software projects. 
These prior works give rise to a blossom of SDP studies that are commonly 
referred to as Just-In-Time Software Defect Prediction (JIT-SDP).

Pursuing the best performing models, prior JIT-SDP studies leverage and
experiment on a variety of modeling and machine learning techniques, such as, 
Logistic Regression~\cite{mockus2000predicting, kamei2012large}, Naive
Bayes~\cite{jiang2013personalized}, Support Vector
Machine~\cite{aversano2007learning, kim2008classifying}, Decision
Tree~\cite{tarvo2013predicting}, and Neural Networks~\cite{yang2015deep}.
Empirical evidences from these works suggest that the 14 software change
metrics can yield high predictive performance and interpretable prediction in
relation to prior software quality studies~\cite{kamei2012large,
khanan2020jitbot, lin2021impact}. In addition, JIT-SDP models using ensemble
models like Random Forest and those using deep learning models like Deep Neural
Networks tend to yield better predictive performance than the
others~\cite{fukushima2014empirical, kamei2016studying, yang2015deep,
hoang2019deepjit, hoang2020cc2vec, xu2021effort}. 

Following the prior studies, we use the 14 software change metrics proposed by
Kamei et al.~\shortcite{kamei2012large} and adopt Deep Neural Networks as
modeling technique to realize a JIT-SDP model in our proposed modeling framework. 

We should note that none of the prior Deep Neural Network models fit in the
modeling framework we propose, an incremental modeling framework embracing
concept-preserving class balancing, combining forecasting and classification,
and addressing concept drift in software changeset data.

\subsection{Incremental JIT-SDP}
Most JIT-SDP works propose and study offline models, i.e., they train the
JIT-SDP models, typically classification models using pre-existing training data
sets and do not consider the problem of updating the models when new software
changesets arrive in the SCMs~\cite{fukushima2014empirical, kamei2016studying,
yang2015deep, hoang2019deepjit, hoang2020cc2vec, xu2021effort}. However,
empirical evidence shows that incremental or online learning is a more
realistic learning setting for JIT-SDP~\cite{tan2015online, mcintosh2017fix,
cabral2019class, bennin2020revisiting, tabassum2020investigation}. Incremental
or online learning models are those that learn from and adapt to new training
data instances. Although there is no consistent definition, we adopt the view
that in incremental learning we update a model when a chunk of (or $N$, $N \ge
1$) training data instances arrive and in online learning we do for each new
training instance~\cite{zhang2019incremental}.  In this view, we consider
online learning is a special case of incremental learning where $N = 1$.

Tan et al.~\shortcite{tan2015online} argue that development characteristics,
e.g., development tasks, developer experience, and programming styles may
change significantly if over a long period of time, as a result, the training
data set may be too old to train accurate prediction models for the test data
set.  This is one of the factors that lead to their proposal of in effect an
incremental learning JIT-SDP model. In their experiments using 6 open-source
projects and 1 proprietary project, although not consistently cross board,
there is an improvement of predictive performance (measured by F1 and
Precision) on most of the 7 software projects when the models get updated.

The very assumption that JIT-SDP models work is that past defect-inducing
changes are similar to future ones. This assumption may no long be valid as the
training data set is too old when compared to the test data set. To investigate
this problem, McIntosh and Kamei~\shortcite{mcintosh2017fix} select two large
rapidly evolving software projects, QT and OpenStack.  They show that the
characteristics of software change metrics of defect-inducing software
changesets fluctuate over time and as a result, an aged JIT-SDP model, a model
trained using the dated software changesets several months or one year old has
indeed degraded predictive performance when compared to a fresh one, the one
trained using the software changesets in the immediate past. Via five
open-source software projects, Bennin et al.\shortcite{bennin2020revisiting}
assert that the predictive performance of a JIT-SDP predictor constructed with
data from historical versions may degrade over time due to the change of the
data distribution, which we consider a confirmation of the findings of McIntosh
and Kamei~\shortcite{mcintosh2017fix}.

A generalization of the fluctuation of the characteristics of software
changesets over time is concept drift~\cite{cabral2019class,
bennin2020revisiting, tabassum2020investigation}, i.e., the change over time of the
relation between the input (i.e., the independent) variables and the target
(i.e., the dependent variable)~\cite{gama2014survey}.  For probability models,
this is the longitudinal variation of the joint probability distribution of the
problem over time~\cite{gama2014survey}. For a software project, clean software
changesets are generally far more than defective changesets, i.e., software
changeset data sets are class imbalanced with regard to their defectiveness
status. Cabral et al.~\shortcite{cabral2019class} considers the evolution of a
software project's class imbalance status as a source of concept drift. They
propose an online JIT-SDP model to adapt to the class imbalance status.
Addressing the problem of insufficient initial training data set in online
JIT-SDP models, Tabassum et al.~\shortcite{tabassum2020investigation} propose
three new approaches to adapt to concept drift in the cross-project setting.

Complement to the prior incremental and online learning studies, our work leads
to two new modeling concepts for incremental JIT-SDP models. First, we
characterize concept drift using principal curves, a summary of joint
probability distribution of the underlying problem. The concept drift
characterized by principal curves represents a drift of the underlying
data distribution, and is beyond the class
imbalance evolution. Second, we propose an incremental learning framework that
combine both forecast and classification components. Both of these are not
present in the prior works.

\subsection{Principal Curves}
\label{sec:related:pc}
Principal curves are ``self-consistent'' smooth curves that pass through the
``middle'' of a multidimensional (or $d$-dimensional) probability
distribution~\cite{hastie1989principal, kegl2000learning}, where
self-consistency means that each point of a principal curve is the average of
all points under the probability distribution that project
there~\cite{kegl2000learning}. Principal curves are thought as a nonlinear
generalization of Principal Component Analysis and have found applications in
areas like dimensional reduction and feature extraction. Recent examples
include dimensional reduction for data streams~\cite{li2021sequential} in
sequential learning setting, feature extraction for clustering from functional
data~\cite{wu2021functional}, and image processing for object
detection~\cite{parmiggiani2019image}.

In practice, we possess a data set, represented by $d$-dimensional vectors,
however, without knowing the underlying probability distribution of the
problem.  By treating the vectors of the data set as the samples of the
underlying distribution, we can apply a non-parametric statistical method to
estimate the principal curve for the data set, and the principal curve
characterizes the underlying distribution of the
data~\cite{hastie1989principal, kegl2000learning}. The principal curve becomes
a non-linear summary of the data and its underlying
distribution~\cite{hastie1989principal, kegl2000learning}.

Our motivation of adopting principal curves as a non-linear summary of the
software changeset data and as the means to implement an over-sampling class
imbalancing process technique comes from Mao et al.~\cite{mao2015online}.  For
the class imbalance problem, Mao et al. argue that over-sampling technique is a
favorable method than under-sampling technique because over-sampling does not
drop data samples that may contain useful information. A typical over-sampling
technique is synthetic minority over-sampling technique (SMOTE).  However,
SMOTE interpolates virtual samples for minority class and due to randomness
there is no guarantee that the new minor class samples generated by SMOTE
satisfy the distribution of the raw data~\cite{mao2015online}. To address this
problem, Mao et al. propose to use the distance to the principal curves of the
raw data to determine  the acceptance of SMOTE generated data instances.  Following
Mao et al., we propose a simple algorithm that combines SMOTE and
principal curves and the algorithm ensures that the minority class data set
after SMOTE has a similar principal curve to the  original raw data.

In the context of JIT-SDP, none of the prior works attempt to characterize the
distribution of the software changeset data, to represent concept drifts in the
software changeset data using principle curves, and to preserve the
``concepts'' in the software changeset data, i.e., the underlying distribution
of the data when balancing the software changeset data for training a JIT-SDP
model.

\section{Data Preparation for Online learning}
\label{sec:data}
\begin{figure}[!htbp]
	\centering
	\begin{adjustbox}{max width=1\columnwidth}
 	\begin{tikzpicture}
 		\tikzset{
			node distance=0.125in and 0.30in,
 			every node/.style = {
 				text=black,
 				fill=white,
 				font=\footnotesize\sffamily
 			},
 			io/.style = {
 				rectangle,
				minimum width=0.7in,
 			},
			datasource/.style = {
				rectangle,
				dashed,
				draw,
				minimum width=0.8in,
		  },
			tallprocess/.style = {
				rectangle,
				solid,
				draw,
				minimum width=0.7in,
			},
			process/.style = {
				rectangle,
				solid,
				draw,
				minimum width=1.7in,
			},
			align=center,
			wrapper/.style = {
				rectangle,
				dotted,
				draw,
				draw opacity = 1,
				fill opacity = 0,
			},
			desc/.style = {
				rectangle,
				font=\itshape
			}
 		}
 
		\node(ours)[io]{Select OSS Projects};
		\node(repo)[datasource, below=of ours, xshift=-0.8in]{Github Repository};
		\node(jira)[datasource, below=of ours, xshift=0.7in]{Jira ITS};
		\node(getchange)[tallprocess, below=of repo, xshift=-0.5in]{Commit Diffs};
		\node(getlog)[tallprocess, below=of repo, xshift=0.5in]{Commit Messages};
		\node(getmetric)[tallprocess, below=of getchange]{Change Metrics};
		\node(getlabel)[tallprocess, below=of getlog, xshift=0.4in]{Run SZZ};
		\node(getireport)[tallprocess, below=of jira]{Issue Reports};
		\node(changeset)[datasource, below=of getmetric, xshift=0.5in]{Labeled Changesets};
		\node(part)[process, below=of changeset]{Order \& Partition};
		\node(log)[process, below=of part]{Logarithmic Transformation};
		\node(fs)[process, below=of log]{Feature Selection or Removal};
		\node(final)[datasource, below=of fs]{Processed Changesets};

		\draw[-stealth](ours) -- (repo);
		\draw[-stealth](ours) -- (jira);
		\draw[-stealth](repo) -- (getchange);
		\draw[-stealth](repo) -- (getlog);
		\draw[-stealth](jira) -- (getireport);
		\draw[-stealth](getchange) -- (getmetric);
		\draw[-stealth](getlog) -- (getlabel);
		\draw[-stealth](getireport) -- (getlabel);
		\draw[-stealth](getmetric) -- (changeset);
		\draw[-stealth](getlabel) -- (changeset);
		\draw[-stealth](changeset) -- (part);
		\draw[-stealth](part) -- (log);
		\draw[-stealth](log) -- (fs);
		\draw[-stealth](fs) -- (final);

		\node(ourwrapdesc)[desc, left=of ours, xshift=-0.45in]{%
			Group 2};
		\node(ourwrap)[wrapper, fit=(ourwrapdesc) (ours) (getmetric) (jira)]{};

		\node(ppdesc)[desc, right=of changeset]{%
			Groups 1 \& 2};
		\node(ppwrap)[wrapper, fit=(ppdesc) (changeset) (part) (fs) (final)]{};

 	\end{tikzpicture}
\end{adjustbox}
	\caption{Flow of data acquisition and preprocessing. Top part of the flow is only
	applicable to Group 2 projects, i.e., the bottom group in Table~\ref{tab:datasets}.}
	\label{fig:preprocessing}
\end{figure}
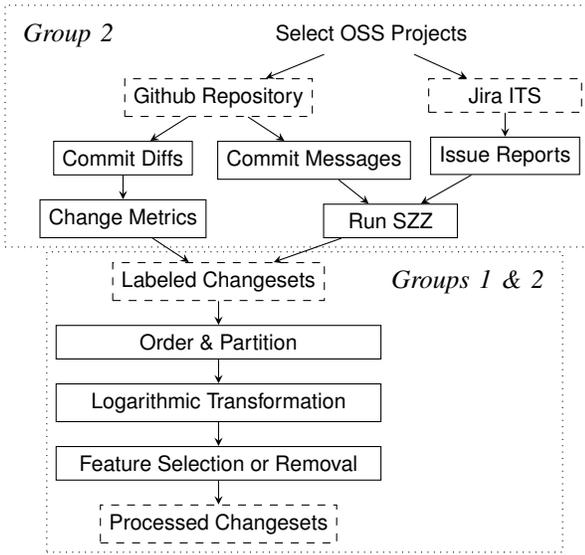

To answer the research questions, we use the changesets and their
defectiveness inducing status from the \numprojs open-source software projects. We
describe the projects, data acquisition, and  preprocessing of the data in the
following.  Figure~\ref{fig:preprocessing} illustrates the steps of data
acquisition and preprocessing.

\subsection{Dataset}
Table~\ref{tab:datasets} is a summary of the \numprojs projects. We assign a
short code to each software project and refer to the software projects using
their short codes in the tables in the rest of this article.  
The data sets are in two groups. The first group (called group 1 thereafter) is
the data set of the first 4 software projects from Eclipse JDT to PostgreSQL in
the table.  This data set is available from Kamei et
al.~\shortcite{kamei2012large}. The changesets of this data set are logical
transactions from CVS or Subversion repositories where a logical transaction
are related commits in a short time window~\cite{kamei2012large}.

The second group (called group 2 thereafter) is the rest of the \numapacheprojs
open source projects by the Apache Software Foundation.  Two reasons lead us to
select these software projects. First, they are often the subjects of the
studies on software quality and defect prediction~\cite{just2014defects4j}.
Second, these software projects use an issue tracking system (ITS) to manage
their development process and require developers to reference the ITS issues in
their Git commits. As we discuss below, this helps us identify defect-inducing
changesets, a changeset that induces a defect in the software
code~\cite{kamei2012large}. Defect-inducing changesets are often called
bug-inducing changesets (BICs)~\cite{rosa2021evaluating} or sometimes, perhaps
more precisely, fix-inducing changesets (FICs)~\cite{mcintosh2017fix} as we
shall show below that the identification begins from a changeset that fixes
a defect.

\begin{table*}
	\centering
	\caption{\label{tab:datasets}Description of Studied Software Projects}
	\begin{adjustbox}{max width=\textwidth}
	\begin{tabular}{l l r r r l l p{0.28\textwidth}}
		\toprule

		\makecell{Project}  
			& \makecell{Short\\Name}
			& \makecell{\# of \\Changesets} 
			& \makecell{\# of \\BICs}
			& \makecell{Clean-to-Defect\\Changeset Ratio}
			& \makecell{Period of \\Changesets} 
			& \makecell{Programming\\Language} 
			& \makecell{Application\\Domain}
		\\
		\midrule
		Eclipse JDT 
			& JDT
			& 35,386 
			& 5,089
			& 5.6:1
			& 05/2001--12/2007 &Java &  Java development tools \\
		Eclipse Platform 
			& PLA
			& 64,250 
			& 9,452
			& 5.8:1
			& 05/2001--12/2007 & Java & Integrated Development Environment \\
		Mozilla Browser 
			& MOZ
			& 98,275 
			& 5,149
			& 18.0:1
			& 01/2000--12/2006 & C++ & Web browser \\
		PostgreSQL 
			& POS
			& 20,431 
			& 5,119
			& 3.0:1
			& 07/1996--05/2010 & C  & database management system \\
		\midrule

		ActiveMQ 
			& AMQ
			& 12,349
			& 4,231
			& 1.9:1
			& 12/2005--08/2021 & Java & message broker \\
		Camel 
			& CAM
			& 59,451
			& 7,754
			& 6.7:1
			& 03/2007--08/2021 & Java & message-oriented middleware \\
		Hadoop 
			& HDP
			& 56,065
			&  2,566
			& 20.9:1
			& 05/2009--08/2021 & Java & map-reduce framework\\ 
		HBase 
			& HBA
			& 35,022
			&  6,116
			& 4.7:1
			& 04/2007--08/2021 & Java & database \\
		Mahout 
			& MAH
			& 4,578
			& 1,391
			& 2.3:1
			& 01/2008--06/2021& Java & machine learning framework \\
		OpenJPA 
			& JPA
			& 7,344
			& 1,671
			& 3.4:1
			& 05/2006--06/2021 & Java  & Java persistence library\\	
		\bottomrule
	\end{tabular}
	\end{adjustbox}
\end{table*}

\subsubsection{Changesets Acquisition and Feature Extraction}
Kamei et al.~\shortcite{kamei2012large} study 14 software change metrics in
their JIT-SDP models.  A blossom of JIT-SDP studies ensure and give these
metrics a thorough examination.  Table~\ref{tab:changemetrics} is the list of
the software change metrics and their brief description.

\begin{table}[!htbp]
	\centering
	\caption{Software Change Metrics~\cite{kamei2012large}}
	\label{tab:changemetrics}
	\begin{tabular}{l l p{2.00in}}
		\toprule
		Category & Metric & Description \\
		\midrule
		\multirow{4}{*}{Diffusion}
			& \metric{NS} & Number of modified subsystems\\
		 	& \metric{ND} & Number of modified directories\\
		 	& \metric{NF} & Number of modified files\\
			& \metric{Entropy} & Distribution of modified code across each file \\
		 \midrule
		\multirow{3}{*}{Size}
			& \metric{LA} & Lines of code added\\
		 	& \metric{LD} & Lines of code deleted\\
		 	& \metric{LT} & Lines of code in a file before the change \\
			& \metric{Churn} & Size of the change, i.e., \metric{LA} + \metric{LD} \\
		\midrule
		Purpose & \metric{FIX} & Whether the change is a defect fix \\
		\midrule
		\multirow{3}{*}{History}
			& \metric{NDEV} & The number of developers that changed the modified files \\
		 	& \metric{AGE} & The average time interval between the last and current change \\
		 	& \metric{NUC} & The number of unique changes to the modified files \\
		\multirow{3}{*}{Experience}
			& \metric{EXP} & Developer experience \\
		 	& \metric{REXP} & Recent developer experience \\
		 	& \metric{SEXP} & Developer experience on a sub-system \\
		\bottomrule
	\end{tabular}
\end{table}

For group 1 software projects, Kamei et al. make available the software change
metrics in their shared data set.  For group 2 software projects, the authors
of this article
extract commits from the Git repositories of these projects on Github and treat
each commit as a changeset, and compute the 14 software change metrics
following Fan et al.~\shortcite{fan2019impact}, the process of which is in
Figure~\ref{fig:preprocessing}.

\subsubsection{Identifying Defect-Inducing Changesets}
Following Kamei et al.~\shortcite{kamei2012large}, we identify BICs 
for group 2.  The projects in
this group use Jira as their ITS. Via Jira's API, we first {\em retrieve issues
reports}
whose ``\code{issuetype}'' is ``\code{Bug}'', whose status is ``\code{Resolved}''
or ``\code{Closed}'', and whose resolution is ``\code{Fixed}''. Since
developers reference issue numbers in their Git commit messages as part of the
development process, we retrieve the commits by looking for the issue numbers
in the commit messages. These commits are fixing commits or fixing changesets
(FCs), a commit or a changeset that fixes a known defect. 
With the FCs and the commit history of a project
as input, 
we run the SZZ algorithm to identify BICs as illustrated in
Figure~\ref{fig:preprocessing}. The SZZ implementation is
the one in PyDriller~\cite{spadini2018pydriller}. 
It filters out cosmetic changes and uses annotation graphs to identify
BICs~\cite{spadini2018pydriller} and is based on B-SZZ, the original
SZZ implementation~\cite{sliwerski2005changes}. A recent comprehensive study show that this
implementation, although not the best,  has an overall reasonable ability to identify
BICs~\cite{rosa2021evaluating}. Furthermore, a study of SZZ implementations
shows that B-SZZ and MA-SZZ are
the SZZ implementations that are unlikely to
cause a considerable performance reduction due to misidentified
BICs in JIT-SDP~\cite{fan2019impact}. Our own sampling of identified
BIC's in the Group 2 projects indicate that PyDriller's SZZ implementation
identifies most commits that have defect-introducing code.

\subsection{Preprocessing}
This work is to design and to evaluate an {\em incremental} learning framework
for JIT-SDP. For this, as shown in Figure~\ref{fig:preprocessing}, we order the
software changesets according to the order in which they arrive in the SCMs,
and then partition the changeset into multiple segments.

Software metrics data are often skewed and sometimes suffer from collinearity
and multicollinearity problems~\cite{kamei2012large}. We follow established
practice in JIT-SDP studies~\cite{kamei2012large} to process the software
metrics data. First, we apply a {\em logarithmic transformation} to the metrics
(excluding \metric{FIX}), i.e., apply $\log (x + 1)$ to the metrics where $x$
represents the value of a metric~\cite{kamei2012large}. 
Second, we address
the collinearity and multicollinearity problems by removing highly dependent
metrics~\cite{kamei2012large}. The removed metrics are project-dependent, and
removed metrics are generally a subset of metrics \metric{ND}, \metric{REXP},
\metric{ENTROPY}, \metric{NUC}, and \metric{FIX}.

\section{Exploring Relationship of Software Changesets}
\label{sec:relation}

We answer RQ.1 in this section. We empirically demonstrate that
\begin{inparaenum} 
\item software changesets are not independent events and
exhibit temporal interdependent relationship, and 
\item we can harness the interdependent
relationship to predict defective changes using an autoregressive forecast
model better than a baseline random prediction model.  
\end{inparaenum}

\subsection{Software Changeset Relationship}

We maintain all software changesets in the order of arrival in their SCMs. For
the ordered changesets, we extract ordered tuples of software changesets, and
compute the conditional and joint probabilities of the tuples. To make this
clear, let us denote the sequence of software changesets in their arrival order
is $(c_1, c_2, \ldots, c_i, c_{i+1}, $\ldots$, c_n)$ where $c_1$ is the oldest
and $c_n$ the most recent. For this sequence, we
have pairs of software changesets, e.g., $(c_1, c_2)$, $(c_2, c_3)$ , $\ldots$,
$(c_i, c_{i+1})$, $\ldots$, $(c_{n_1}, c_n)$, or triplets of software
changesets, $(c_1, c_2, c_3$, $(c_2, c_3, c_4)$, $\ldots$, $(c_i, c_{i+1},
c_{i+2})$, $\ldots$, and $(c_{n-2}, c_{n-1}, c_n)$.  We treat the software
change metrics and label of the software changesets as random variables. For
instance, the label is a random variable that takes value in $\{0, 1\}$ where 0
means clean and 1 defect-inducing.  We then estimate conditional probabilities
and marginal probabilities of interested random variables by counting. 

To ascertain whether software changesets are independent, we conduct a
chi-square test of independence~\cite{franke2012chi}. For this we begin with
building contingency tables consisting of counts of patterns $(0, 0)$, $(0, 1)$,
$(1, 0)$, and $(1, 1)$ of changeset pairs of $(c_{t-1}, c_t)$. In addition,
we define the concept of intersecting changesets. Changeset $c_t$ is an
intersecting changeset when it modifies one or more files that $c_{t-1}$
also modifies. We analyze Group 2 software projects whose code repositories
are available to us by first excluding
merge commits and then counting intersecting commits. Table~\ref{tab:relation:labelpairs}
is a summary of these two analyses, which shows that we safely reject the
null hypothesis that changesets are independent and one possible source of this
phenomenon can be that there exist a significant portion of intersecting changesets.

\begin{table}[!htbp]
	\centering
	\caption{Chi-Squared Test of Independence and Statistics of Intersecting Changesets}
	\label{tab:relation:labelpairs}
	\begin{tabular}{c c r r}
		\toprule
		Project & Intersected Changesets\% & $\chi^2$ & p-value \\
		\midrule
		AMQ & 11.6\% & 215.3 & $\ll 1.0e-10$ \\
		CAM & 19.8\% & 1107.9 & $\ll 1.0e-10$ \\
		HDP & 22.3\% &34.5 & $4.1e-09$ \\
		HBA & 20.1\% & 383.3 & $\ll 1.0e-10$ \\
		MAH & 34.3\% & 170.0 & $\ll 1.0e-10$ \\
		JPA & 15.2\% &240.4 & $\ll 1.0e-10$ \\
		\bottomrule
	\end{tabular}
\end{table}

Furthermore, we examine the patterns that changesets may exhibit.
For this, we examine the triplets of software changesets and
estimate the joint probabilities of the triplets with regard to the labels
of the changesets. Table~\ref{tab:relation:triplets} is the summary of the
joint probabilities of the software changeset triplets with regard to the
labels, e.g., $P_{010} = P((0, 1, 0))$, the probability of the
triplets whose labels are clean,
defect-inducing, and clean. We also computed the confidence interval
for confidence level 95\% following the Central Limit Theorem. For all triplets
of all software projects, confidence interval half width rarely exceeds 10\%
of the estimated probability value in the table.
	The joint probabilities indicate
some triplets almost never appear in these software projects and while
some are significantly more frequently than the others. 
	
\begin{table}[!htbp]
	\centering
	\caption{Probabilities of software changeset triplets with regard to the
	labels}
	\label{tab:relation:triplets}
		\begin{adjustbox}{max width=\columnwidth}
	\begin{tabular}{l r r r r r r r r}
		\toprule
		Project  & $P_{000}$ & $P_{001}$ & $P_{010}$ & $P_{100}$ & $P_{011}$ & $P_{101}$ & $P_{110}$ & $P_{111}$ \\
		\midrule
		JDT  & 0.65 & 0.10 & 0.00 & 0.10 & 0.11 & 0.02 & 0.00 & 0.03\\
		PLA  & 0.65 & 0.09 & 0.00 & 0.09 & 0.11 & 0.02 & 0.00 & 0.03\\
		MOZ  & 0.86 & 0.04 & 0.00 & 0.04 & 0.04 & 0.00 & 0.00 & 0.01\\
		POS  & 0.46 & 0.12 & 0.00 & 0.12 & 0.17 & 0.04 & 0.00 & 0.08\\
		\midrule
		AMQ  & 0.34 & 0.12 & 0.00 & 0.12 & 0.20 & 0.08 & 0.00 & 0.14 \\
		CAM  & 0.69 & 0.08 & 0.00 & 0.08 & 0.10 & 0.02 & 0.00 & 0.03 \\
		HDP  & 0.87 & 0.04 & 0.00 &	0.04 & 0.04 & 0.00 & 0.00 &	0.00 \\
		HBA  & 0.59 & 0.10 & 0.00 & 0.10 & 0.13 & 0.03 & 0.00 & 0.05 \\
		MAH  & 0.41 & 0.11 & 0.00 & 0.11 & 0.17 & 0.06 & 0.00 & 0.13 \\
		JPA  & 0.52 & 0.11 & 0.00 & 0.11 & 0.14 & 0.04 & 0.00 & 0.08 \\
		\bottomrule
	\end{tabular}
		\end{adjustbox}
\end{table}

\subsection{Harnessing Software Changeset Relationship}
\label{sec:relation:auto}
The simple exploration in the above shows software changesets are unlikely 
	independent and the sequences of changesets likely exhibit patterns. The
	question that follows is, can we leverage on this understanding of the
	interdependent relationship among software changesets to aid software defect
	prediction? 

To answer this question, we build two models, a baseline model and an
autoregressive model to forecast defectiveness status of software
changesets. 

The baseline model assumes that we have no information at all
about a software changeset. To predict the changeset to be defect-inducing, it
is akin to tossing a fair coin, and when the coin lands on the head, we
declare the changeset is defect-inducing; clean otherwise. Kamei et
al.~\shortcite{kamei2012large} use this baseline model to evaluate their
JIT-SDP models. 

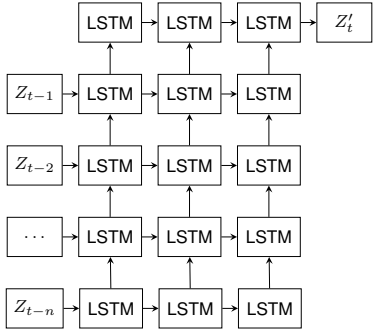
\begin{figure}[!htbp]
	\centering
	\begin{adjustbox}{max width=0.55\columnwidth}
	\begin{tikzpicture}
		\tikzset{
			node distance=0.20in and 0.10in,
			every node/.style={
					text=black,
					fill=white,
					font=\footnotesize\sffamily
			},
			every path/.style={
					color=black,
			},
			align=center,
			io/.style={
				rectangle,
				draw,
				minimum width=0.35in,
				minimum height=0.25in
			},
			lstm/.style={
				rectangle,
				draw,
				minimum width=0.40in,
				minimum height=0.25in
			},
			flow/.style={
				-stealth
			},
		}

		\node(emptyt0)[io, draw opacity=0]{};

		\node(ft1)[io, below=of emptyt0]{$Z_{t-1}$};
		\node(ft2)[io, below=of ft1]{$Z_{t-2}$};
		\node(fto)[io, below=of ft2]{$\ldots$};
		\node(ftn)[io, below=of fto]{$Z_{t-n}$};

		\node(lstmt10)[lstm, right=of emptyt0]{LSTM};
		\foreach \i in {1, 2, o, n}
		{
			\node(lstmt1\i)[lstm, right=of ft\i]{LSTM};
		}

		\foreach \i/\j in {n/o, o/2, 2/1, 1/0} {
			\draw[flow](lstmt1\i) -- (lstmt1\j);
		}

		\foreach \i in {1, 2, o, n}
		{
			\draw[flow](ft\i) -- (lstmt1\i);
		}

		\foreach \l/\i in {1/2, 2/3}
		{
			\foreach \j in {0, 1, 2, o, n}
			{
				\node(lstmt\i\j)[lstm, right=of lstmt\l\j]{LSTM};
			}
			\foreach \m/\n in {n/o, o/2, 2/1, 1/0} 
			{
				\draw[flow](lstmt\i\m) -- (lstmt\i\n);
			}
			\foreach \m in {n, o, 2, 1, 0} 
			{
				\draw[flow](lstmt\l\m) -- (lstmt\i\m);
			}
		}

		\node(out)[io, right=of lstmt30]{$Z_{t}^{\prime}$};
		\draw[flow](lstmt30) -- (out);
	\end{tikzpicture}
	\end{adjustbox}
	\caption{An LSTM-based autoregressive model. $X_t$ is the target changeset
		whose defectiveness status $y_t$ is to be determined. However, the 
		trained model
		has no knowledge about $X_t$ and $y_t$. Instead, it completely relies on
		$Z_{t-1}$, $\ldots$, $Z_{t-n}$, the feature vectors and their known
		defectiveness status of the $n$ preceding software changesets, i.e., $Z_i =
		(X_i, y_i)$, $i = t-1, \ldots, t-n$ to predict $Z_{t}^\prime$. For the
		experiments conducted here, we select $n$ as 6, and 
		$t = n+1, n+2, \ldots, N$ where
		N is the number of changesets for a software project.
		} 
	\label{fig:relation:lstm} 
\end{figure}

The autoregressive model is a forecast model. It predicts
whether future unseen software changesets are defect-inducing and by doing
so, it completely relies on the historical software changes and knows nothing
about the future changesets.  Our tool of designing the autoregressive model
is the Long-Short Time Memory (LSTM) neural networks that is shown to have
superior predictive performance when compared to other autoregressive
models, such as ARIMA~\cite{siami2018comparison, yu2019review}.
Figure~\ref{fig:relation:lstm} depicts the design of the autoregressive model.

To get an overall picture whether the historical changesets can be useful to
forecast the defectiveness status of future changesets, we select the Area
Under the Curve of the Receiver Operating Characteristics (AUC-ROC) as the
evaluation criterion. AUC-ROC is a decision threshold free measure. The
theoretical AUC-ROC value of the random model is known, i.e., 0.5 that does not
depend on the class imbalance status of the data~\cite{kamei2012large}.
Table~\ref{tab:relation:lstm} lists the AUC-ROC values of two models for the
\numprojs software projects.  The AUC-ROC values of the baseline models are the
results of 50 repeated runs.  Similar, the AUC-ROC values of the
autoregressive models are also obtained from 50 repeated runs.  It is worth
noting that although the theoretical value of AUC-ROC of the random model is
0.5, there is always an estimation error unless the number of repeated runs
goes to infinity.  The AUC-ROC values show that the autoregressive model
outperforms the baseline models for 8 out of \numprojs software projects, as
indicated by ``+'' in the table. This suggests that for the majority of the
models, an autoregressive model can harness the patterns in the past to make a
forecast.

\begin{table}[!htbp]
	\centering
	\caption{Average AUC-ROC and standard deviation of 50 runs of random
	baseline model and autoregressive model}
	\label{tab:relation:lstm}
	\begin{center}
		\begin{tabular}{l r r c}
			\toprule
			Project & Baseline & Auto-Regressive & Comparison \\
			\midrule
			JDT &(0.500, 0.014) & (0.503, 0.011) & \mplus \\
			POS &(0.500, 0.016) & (0.524, 0.014) & \mplus \\
			MOZ &(0.495, 0.015) & (0.509, 0.011) & \mplus \\
			PLA &(0.499, 0.011) & (0.556, 0.014) & \mplus \\
			\midrule
			AMQ &(0.493, 0.032) & (0.524, 0.013) & \mplus \\
			CAM &(0.495, 0.020) & (0.516, 0.016) & \mplus \\
			HBA &(0.503, 0.018) & (0.519, 0.012) & \mplus \\
			HDP &(0.504, 0.022) & (0.478, 0.016) & \mminus \\
			MAH &(0.510, 0.057) & (0.569, 0.051) & \mplus \\
			JPA &(0.502, 0.032) & (0.482, 0.021) & \mminus \\
			\bottomrule
		\end{tabular}
	\end{center}
\end{table}

\section{Concept Drift}
\label{sec:drift}
We answer RQ.2 in this section. Concept drift is a change of the underlying
distribution of the data over time~\cite{bennin2020revisiting,
wang2018systematic}. 
For JIT-SDP, 
one source of such distribution changes is the evolution of
class imbalance status~\cite{cabral2019class}. Balancing training data set is a
common preprocessing step in JIT-SDP models to address the class imbalance
problem. In online JIT-SDP models, class balancing can be an integral part of
the models that deal with the class imbalance evolution, a type of concept
drift~\cite{cabral2019class, tabassum2020investigation}.  The class balancing
processing, regardless performed as a preprocessing step or as an integral part
of the model, can offset otherwise the problems of learning bias toward the
majority class and poor generalization. However, does the balancing processing
also alter the very ``concept'' that the JIT-SDP models learn to predict defects
from? In this section, we take SMOTE, a frequently used over-sampling balancing
technique as an example and show whether it alters the underlying
distribution of software changeset data.

\subsection{Concept Drift}
\label{sec:cd:condprob}
\begin{figure*}[!htbp]
	\centering
	\subfloat[PLA]{%
    \includegraphics[width=0.19\textwidth]{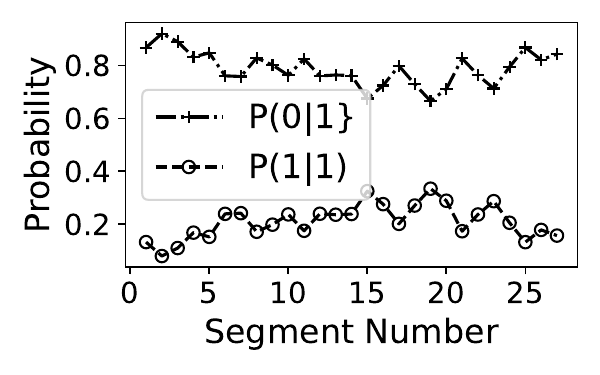}%
    \label{fig:cd:condprob:platform}%
  }
  \hfil
	\subfloat[JDT]{%
    \includegraphics[width=0.19\textwidth]{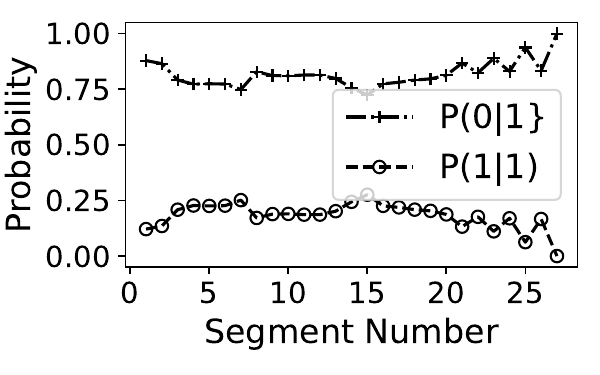}%
    \label{fig:cd:condprob:jdt}%
  }
	\hfil
	\subfloat[MOZ]{%
    \includegraphics[width=0.19\textwidth]{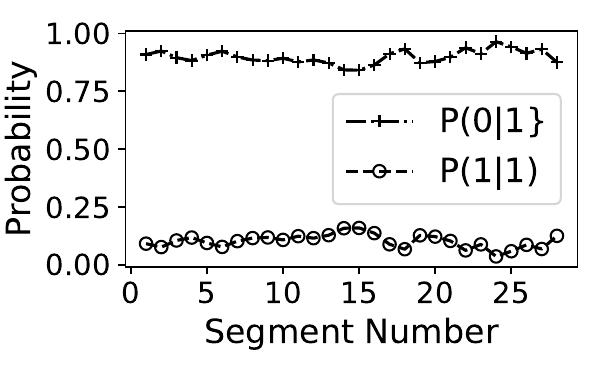}%
    \label{fig:cd:condprob:mozilla}%
  }
	\hfil
	\subfloat[POS]{%
    \includegraphics[width=0.19\textwidth]{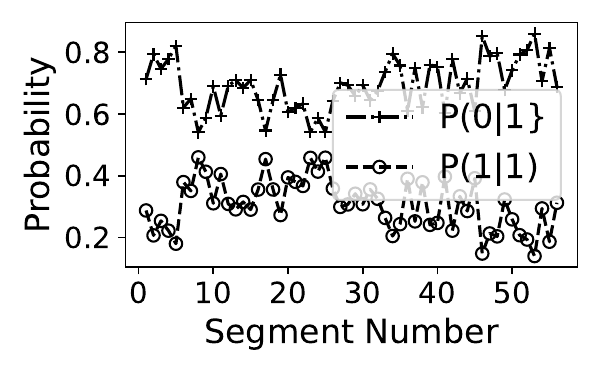}%
    \label{fig:cd:condprob:postgres}%
  }
	\hfil
	\subfloat[AMQ]{%
    \includegraphics[width=0.19\textwidth]{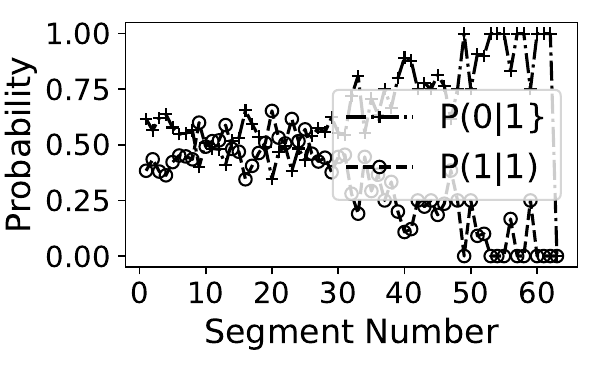}%
    \label{fig:cd:condprob:amq}%
  }

	\subfloat[CAM]{%
    \includegraphics[width=0.19\textwidth]{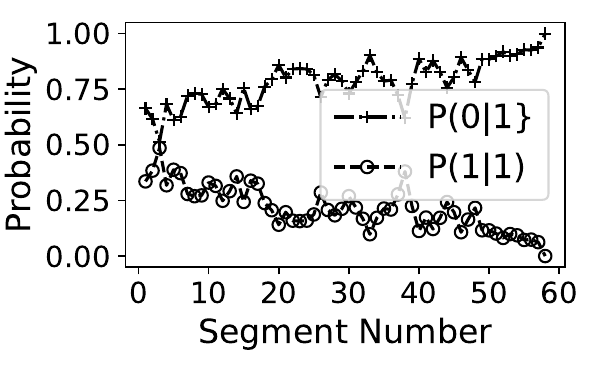}%
    \label{fig:cd:condprob:camel}%
  }
	\hfil
	\subfloat[HDP]{%
    \includegraphics[width=0.19\textwidth]{%
      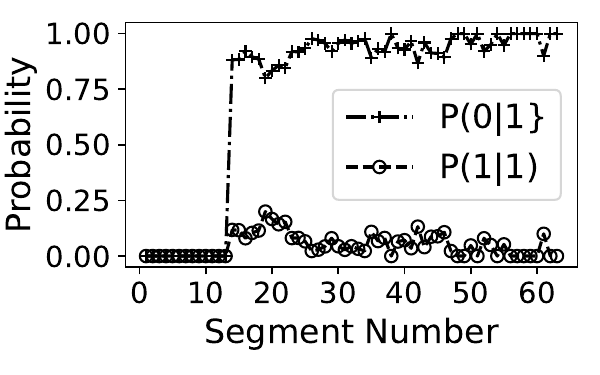}%
    \label{fig:cd:condprob:hadoop}%
  }
	\hfil
	\subfloat[HBA]{%
    \includegraphics[width=0.19\textwidth]{%
      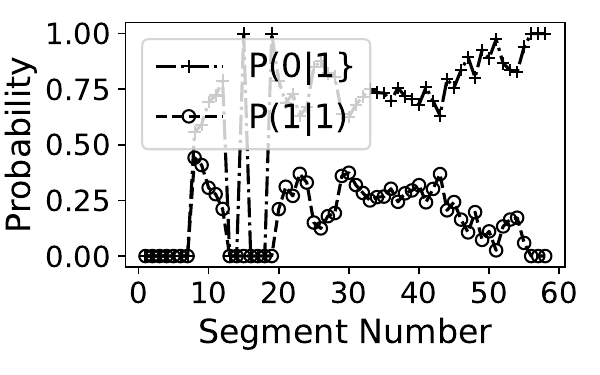}%
    \label{fig:cd:condprob:hbase}%
  }
  \hfil
	\subfloat[MAH]{%
    \includegraphics[width=0.19\textwidth]{%
      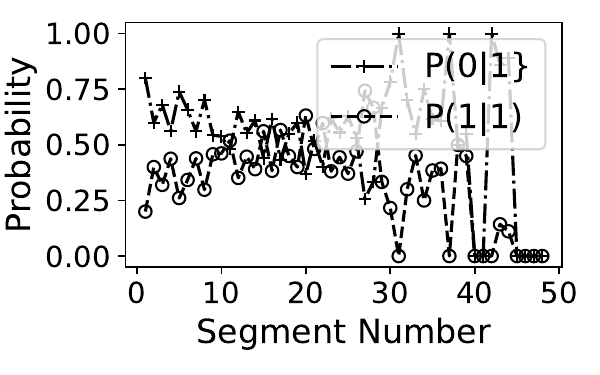}%
    \label{fig:cd:condprob:mahout}%
  }
  \hfil
	\subfloat[JPA]{%
    \includegraphics[width=0.19\textwidth]{%
			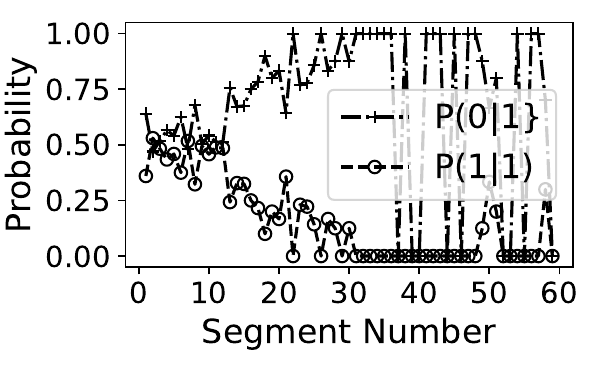}%
    \label{fig:cd:condprob:jpa}%
  }
  \caption{Variations of conditional probabilities $P(1|1)$ and $P(0|1)$ over
  time where each point represents the probability computed in a 3-month
  segment.}
	\label{fig:cd:condprob}
\end{figure*}

Concept drift can be defined as the variation of joint probability distribution of
the problem~\cite{gama2014survey}.  In JIT-SDP, concept drift manifests as
the longitudinal variation of characteristics of software changeset data
features~\cite{mcintosh2017fix, bennin2020revisiting}.  Following the
definition of concept drift, we use probabilities $P(0|1)$ and $P(1|1)$ over
time to illustrate concept drift where $P(0|1)$ is  the conditional probability
at which changeset $X_t$ is clean under the condition the preceding changeset
$X_{t-1}$ is defect-inducing while $P(1|1)$ the conditional probability at
which changeset $X_t$ is also defect inducing under the condition that
changeset $X_{t-1}$ is defect-inducing.
We arrange the software changeset in the chronological order according to their
commit time, divide the data into chronological ordered segments of 3 months,
and estimate the conditional probabilities by counting.  
Figure~\ref{fig:cd:condprob} is
the result. There are several observations.  First, the conditional
probabilities do vary over time. Second, the projects exhibit different
patterns of the longitudinal variations -- the conditional probabilities of
some projects tend to be flat over time, e.g., the Mozilla project while 
some of the others oscillate; the difference between $P(0|1)$ and $P(1|1)$ 
grows large from the early part to the last part of some software projects
while this is not apparent for some of the others.

\subsection{Concept Representation}
\label{sec:cd:pc}
Section~\ref{sec:cd:condprob} provides a snapshot of concept drift in the
software changeset data. However, as the number of random variables of the
problem increases, it becomes intractable to examine all possible combinations
of random variables for their joint, marginal, and conditional probabilities.
We propose to use {\em principal curves} to characterize the underlying
probability distribution of the problem, and use the variation of the principle
curves to depict concept drift. A principal curve is a smooth
one-dimensional curve that provides a non-linear summary of multi-variate
data~\cite{hastie1989principal,rpkgprincurve} as discussed in
Section~\ref{sec:related:pc}. 

\begin{figure*}[!htbp]
	\centering
	\subfloat[Segment 0]{%
 		\includegraphics[width=0.15\textwidth]{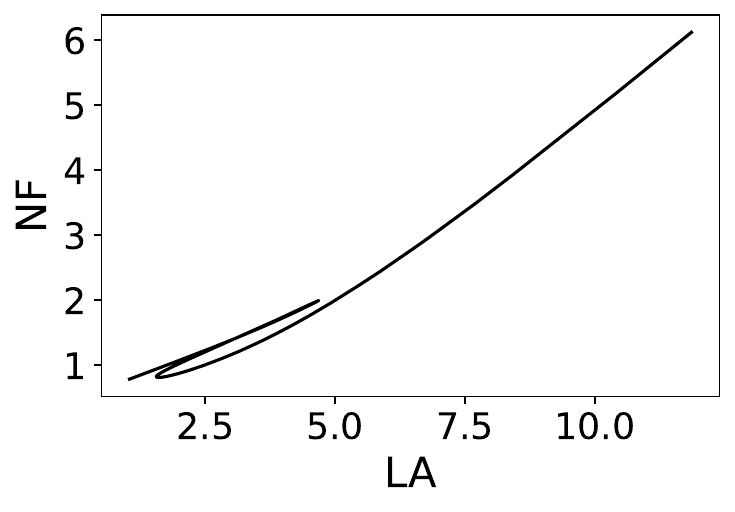}%
		\label{fig:hbasegm1}}
	\hfil
	\subfloat[Segment 1]{%
 		\includegraphics[width=0.15\textwidth]{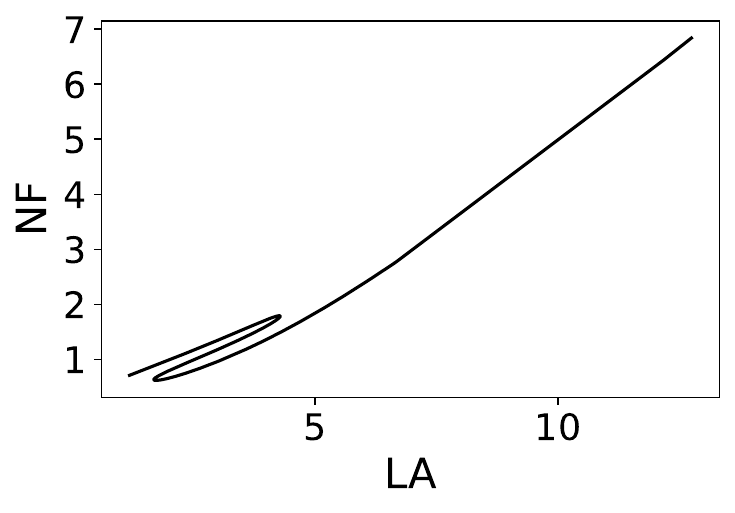}%
		\label{fig:hbasegm2}}
	\hfil
	\subfloat[Segment 2]{%
 		\includegraphics[width=0.15\textwidth]{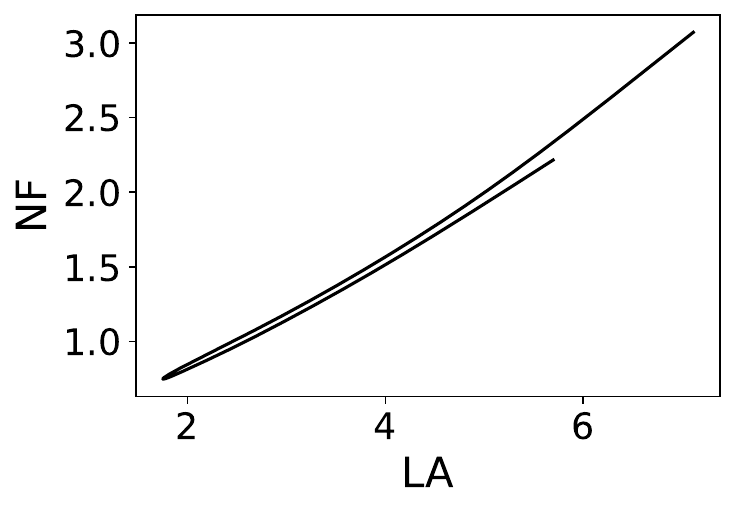}%
		\label{fig:hbasegm3}}
	\hfil
	\subfloat[Segment 3]{%
 		\includegraphics[width=0.15\textwidth]{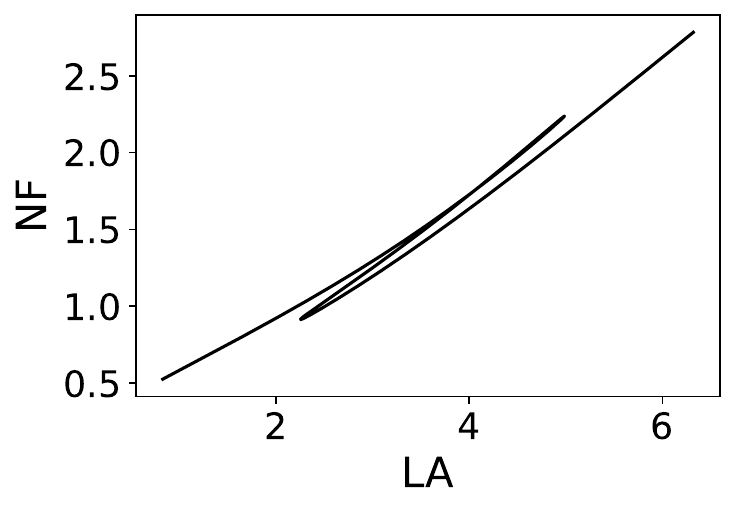}%
		\label{fig:hbasegm4}}
	\hfil
	\subfloat[Segment 4]{%
 		\includegraphics[width=0.15\textwidth]{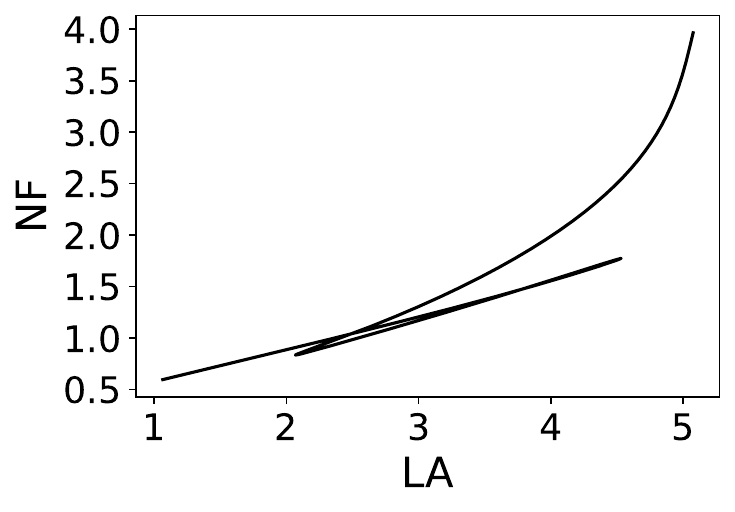}%
		\label{fig:hbasegm5}}
	\subfloat[Similarity]{
		\includegraphics[width=0.15\textwidth]{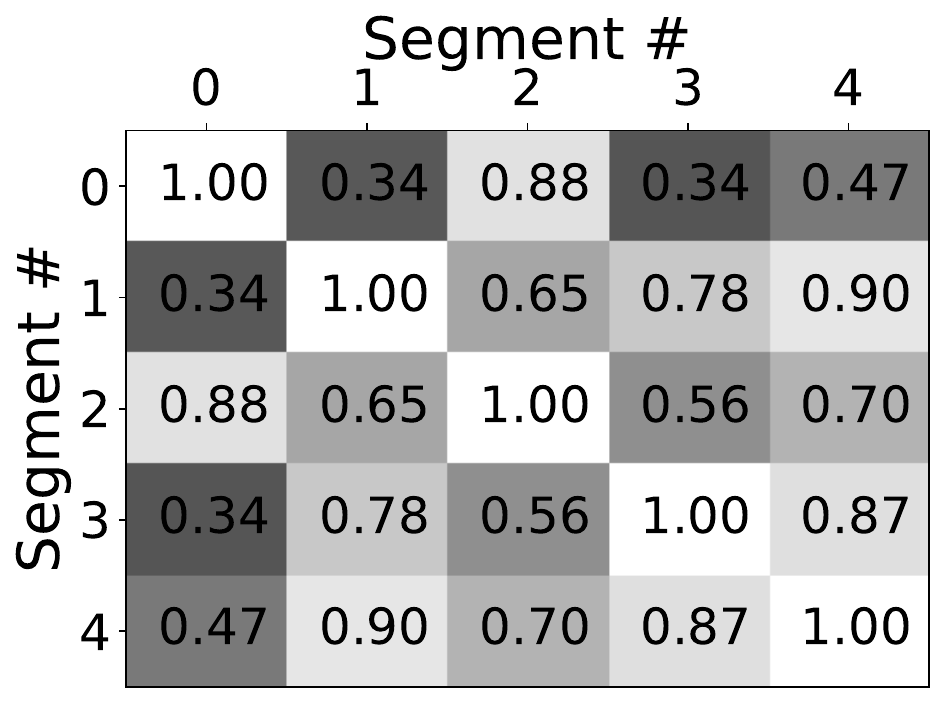}%
		\label{fig:cd:hba:pc:heatmap}
	}
	\caption{Concept drift is visible in different time period of 
	project HBase.
	The principal
	curves are lines in 15-dimensional hyperspace (14 features + 1 label). 
	These figures are the principal curves on a 
	2-dimensional plane of LA vs NF. We choose the LA-NF plane because
	Kamei et al.~\cite{kamei2012large} show that LA and NF are among most important
	contributing features to defect prediction in their study. 
	}
	\label{fig:cd:distpcsegcol}
\end{figure*}

We first divide chronologically ordered software changesets of each software
changeset into 5 segments, each with an identical number of changesets.  We
then compute the principal curves for each segment. For each software project,
we estimate the pair-wise cosine similarity of the principal curves of the
segments.  Figure~\ref{fig:cd:distpcsegcol} plots the principal curves of the 5
chronologically arranged segment of software changeset data of Project
HBase. This example shows that the principle curves of the segments are
clearly different in shape, visually or based on the cosine similarities. Since
the principle curves are the summaries of the underlying probability
distributions, we can argue that there is a concept drift in the software
changeset data of this software project.

\begin{figure}[!htbp]
	\centering
	\subfloat[Clean/Defect Ratio]{
		\includegraphics[width=0.2\textwidth]{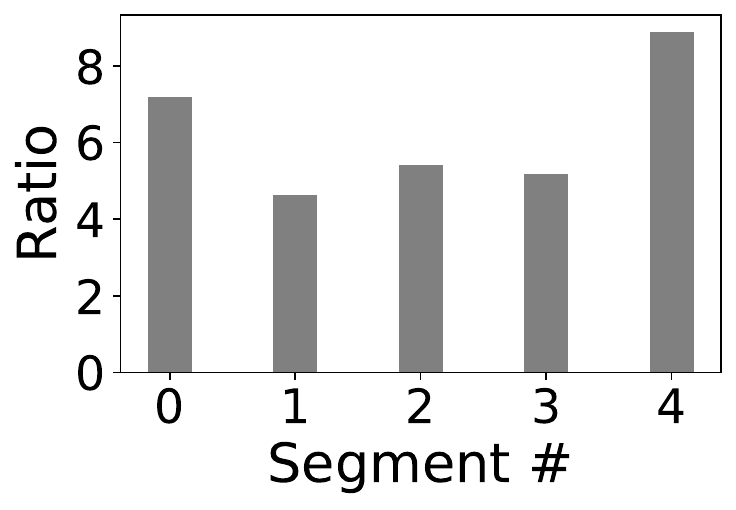}%
		\label{fig:jdtbugratio}}
	\hfil
	\subfloat[Segment 2]{
		\includegraphics[width=0.2\textwidth]{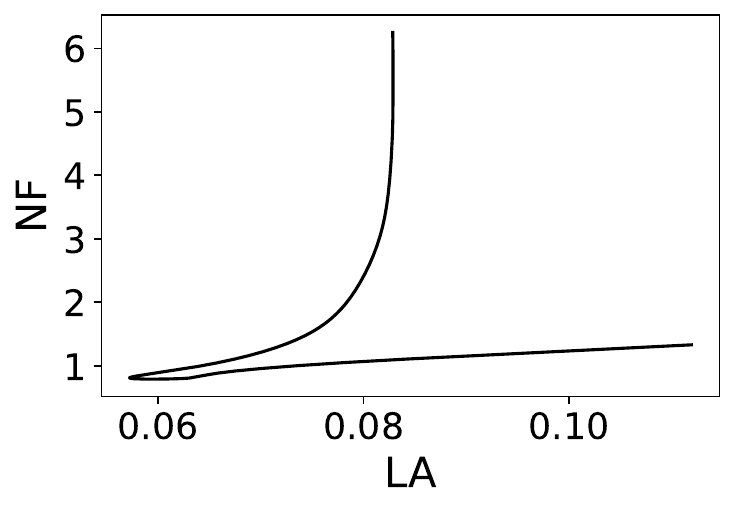}%
		\label{fig:jdtseg2}}
	
	\subfloat[Segment 3]{%
		\includegraphics[width=0.2\textwidth]{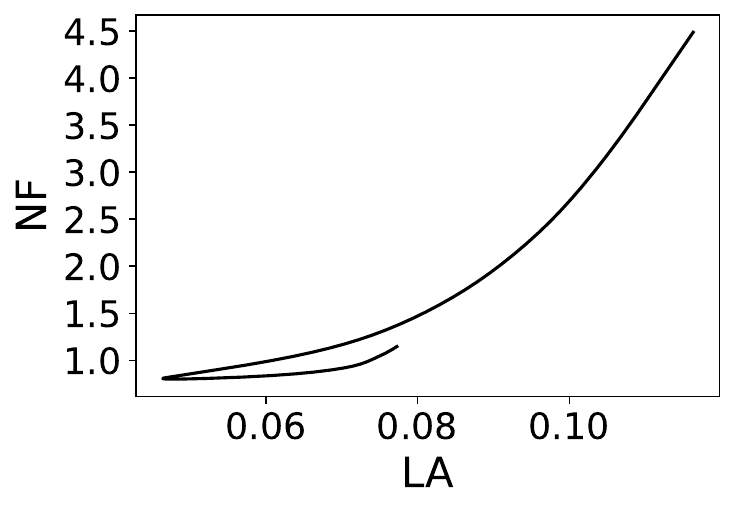}%
		\label{fig:jdtseg3}}
\	\hfil
	\subfloat[Similarity]{%
		\includegraphics[width=0.2\textwidth]{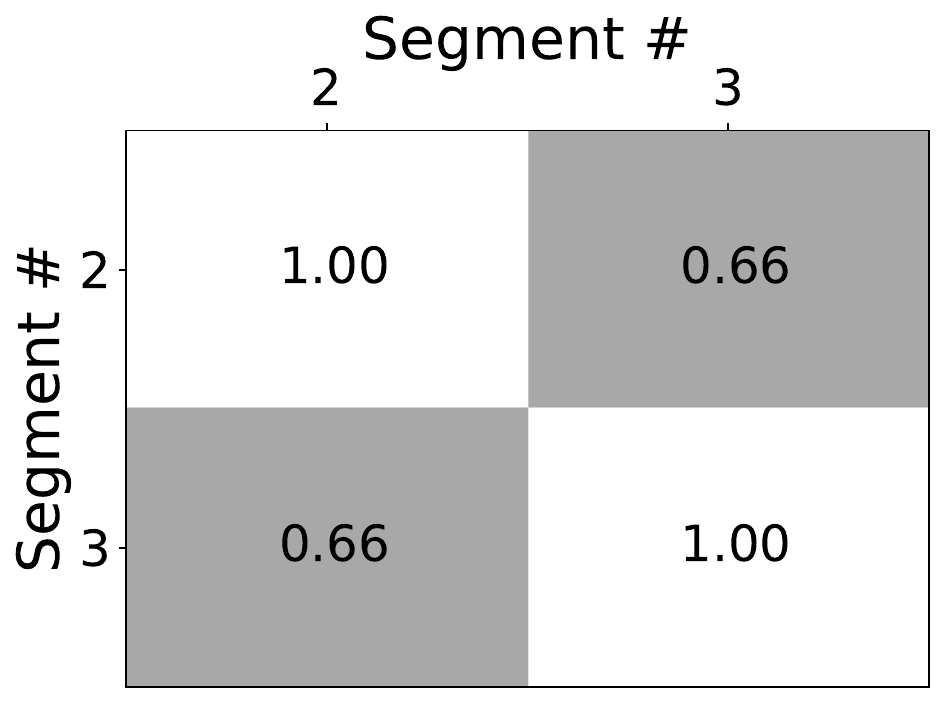}%
		\label{fig:cd:jdt:pc:heatmap:23}}
	\caption{This figure shows that for project Eclipse JDT, concept drift
	visible in the Principal Curves of Segments 2 and 3, but not obvious in the
	class imbalance status of the two segments -- the clean-to-defect changeset
	ratios are almost identical.}

 	\label{fig:cd:pcdriftnoratio}
\end{figure}

\begin{figure}[!htbp]
	\centering
	\subfloat[Clean/Defect Ratio]{
		\includegraphics[width=0.2\textwidth]{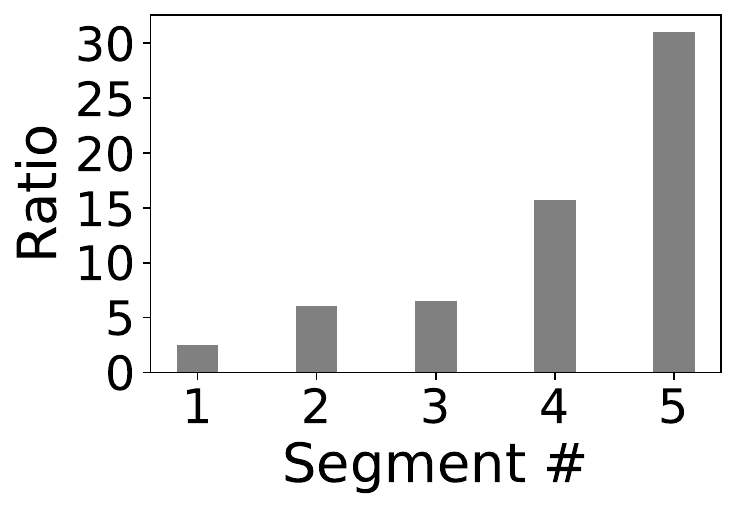}%
		\label{fig:cambugratio}}
	\hfil
	\subfloat[Segment 3]{
		\includegraphics[width=0.2\textwidth]{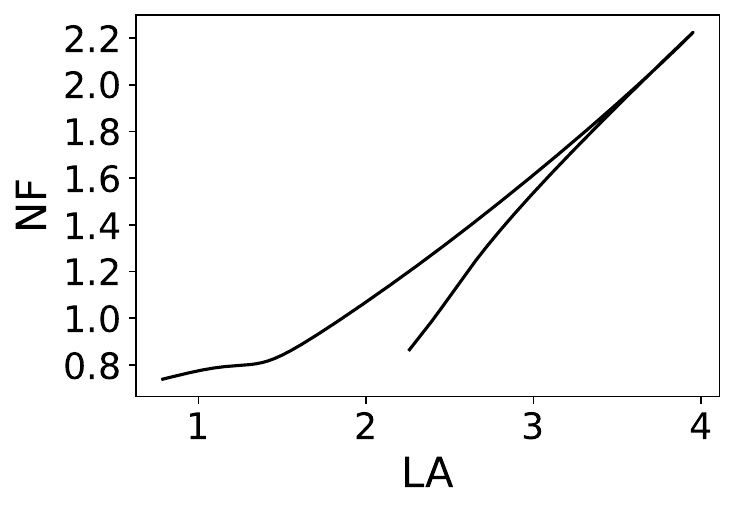}%
		\label{fig:camseg3}}

	\subfloat[Segment 4]{%
		\includegraphics[width=0.2\textwidth]{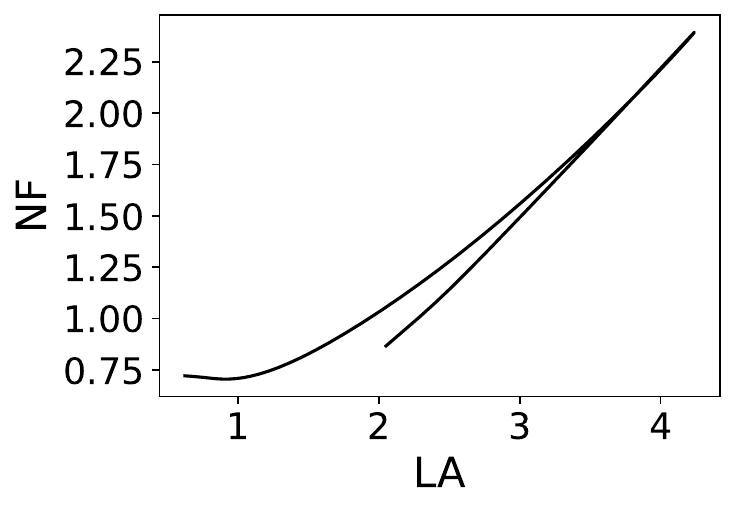}%
		\label{fig:camseg4}}
	\hfil
	\subfloat[Similarity]{%
		\includegraphics[width=0.2\textwidth]{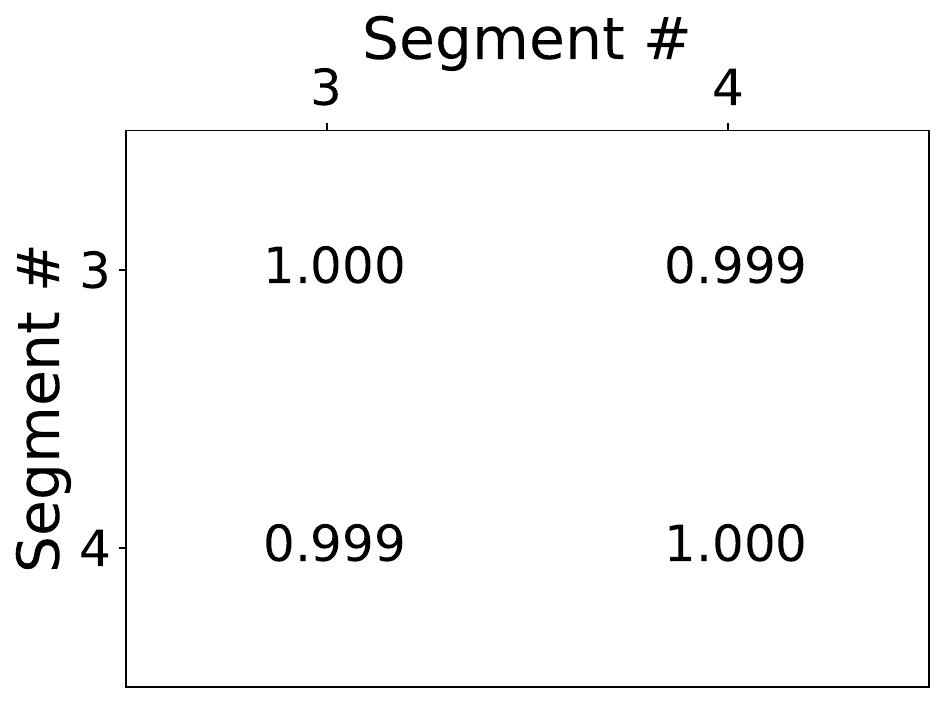}%
		\label{fig:cd:cam:pc:heatmap:34}}
	\caption{This figure shows that for project Camel, concept drift visible
	in the class imbalance status of segments 3 and 4 -- the clean-to-defect 
	changeset ratios visibly differ; however, their principals curves 
	are similar.}
 	\label{fig:cd:noratiopcdrift}
\end{figure}

Prior studies define the class imbalance evolution as a source of concept
drift.  Comparing the class imbalance status with the principle curves of the
software projects, we draw two additional observations. First, there is a
visible difference among the principle curves while there is little variations
of the class imbalance status in some segments of changeset data among some
software projects.  Figure~\ref{fig:cd:pcdriftnoratio} exhibits this
observation.  Second, there is a significant change of class imbalance status
while the principle curves vary little in some segments of data among some
software projects.  Figure~\ref{fig:cd:noratiopcdrift} is an example of the
observation. 

Nevertheless, it is apparent from these observations that the class imbalance
evolution is not a sole source of concept drift.  Using the principle curves,
we characterize a different form of concept drift than the class imbalance
evolution.

\section{Concept-Preserving Class Balancing}
\label{sec:smotepc}

Prior studies consistently observe that JIT-SDP is a class imbalance
problem~\cite{kamei2012large, tan2015online, cabral2019class,
tabassum2020investigation}.  As indicated in Section~\ref{sec:related:pc}, SMOTE
is one of the favorable approaches to address the class imbalance problem;
however, past studies hypothesize that SMOTE may change the underlying
distribution of the data~\cite{mao2015online}.  If this is to occur to the
software changeset data, it may negatively impact predictive performance of
JIT-SDP models that use SMOTE as the method to address the class imbalance
problem. In this section, we show empirically that this hypothesis likely holds
for software changeset data.  Following this, we propose a new class imbalance
processing technique called SMOTE-PC aiming at reducing the impact of SMOTE on
the distribution of the data.

\subsection{Class Imbalance Processing and Data Distribution}

As in Section~\ref{sec:cd:pc}, we divide the preprocessed software changesets
into segments that we we call the ``raw'' changeset data where clean datasets
are far more than defect-inducing changesets. We apply SMOTE to balance the
datasets, i.e., to over-sample the
defect-inducing changesets, we call the union of
the  raw changeset data and the SMOTE generated data set as the SMOTE data set.
Since principal curves are summaries of the underlying distribution of the
data, we compute principle curves for the raw changeset data and the SMOTE data
set.  Two lines in Figure~\ref{fig:smotepc:distpcrq3} are the principal curves
of the raw changeset data and the SMOTE data set of 4 segments of the software
project.  We observe that there is a visible difference between the two
principal curves.  This is an indication that that SMOTE can change the
underlying distribution of the data.

\begin{algorithm}[!htbp]
	\caption{SMOTE-PC}
	\label{alg:smotepc}

	\DontPrintSemicolon
	\LinesNumbered

	\SetKwProg{Function}{Function}{}{end}
	\SetKwFunction{SMOTEPC}{SMOTEPC}
	\SetKwFunction{GetMajorClass}{GetMajorClass}
	\SetKwFunction{GetDataForClass}{GetDataForClass}
	\SetKwFunction{Set}{Set}
	\SetKwFunction{SMOTEPCSample}{SMOTEPCSample}
	\SetKwFunction{StackMatrice}{StackMatrice}
	\SetKwFunction{NumChangesets}{NumChangesets}
	\SetKwFunction{PrincipalCurve}{PrincipalCurve}
	\SetKwFunction{EmptyMatrix}{EmptyMatrix}
	\SetKwFunction{IsSimilar}{IsSimilar}
	\SetKwFunction{SMOTE}{SMOTE}
	\SetKwFunction{RandomSample}{RandomSample}

	\KwIn{$\mtx{D}$: $n \times m$ feature matrix for $n$ software changesets, each
		with $m$ features\\
		\quad\quad\quad $\mtx{L}$: $n \times 1$ label vector for $\mtx{D}$
	}
	\KwOut{$\mtx{D}_{bal}$: balanced feature matrix}

	\Function{\SMOTEPC($\mtx{D}$, $\vect{L}$)}{
		$l_{major}$,$D_{major}$  $\gets$ \GetMajorClass($\mtx{D}$, $\vect{L}$) \\
		$n_{major}$ $\gets$ \NumChangesets{$D_{major}$} \\
		$L_{minor}$ $\gets$ \Set($\vect{L}$)$\setminus l_{major}$ \\
		$\mtx{D}_{bal}$ $\gets$ $\mtx{D}_{major}$ \\
		\ForEach{$l \in L_{minor}$}{
			$\mtx{D}_l$,$n_l$ $\gets$  \GetDataForClass($\mtx{D}$, $\vect{L}$, $l$) \\
			$\mtx{S}_l$ $\gets$ \SMOTEPCSample($\mtx{D}_l$, $n_{major} - n_l$)  \\
			$\mtx{D}_{bal}$ $\gets$ \StackMatrice($\mtx{D}_{bal}$, $\mtx{D}_l$, $\mtx{S}_l$) \\
		}
		\Return $\mtx{D}_{bal}$
	}

	\vskip 1em
	\KwIn{$\mtx{D}$: feature matrix\\
	\quad\quad\quad $k$: $k$ samples to generate}
	\KwOut{$\mtx{S}$: $k \times m$ feature matrix generated from sampling $\mtx{D}$ using SMOTE}

	\Function{\SMOTEPCSample($\mtx{D}$, $k$)}{
		$C$ $\gets$ \PrincipalCurve($\mtx{D}$) \\
		$\mtx{S}$ $\gets$ \EmptyMatrix()\\
		\Repeat{\NumChangesets($\mtx{S}$) $\ge k$}{
			$\mtx{S}^\prime$ $\gets$ \SMOTE($\mtx{D}$)\\
			$\mtx{T}$ $\gets$ \StackMatrice($\mtx{D}$, $\mtx{S}^\prime$) \\
			$C^\prime \gets$ \PrincipalCurve($\mtx{T}$) \\
			\If{\IsSimilar($C$, $C^\prime$)}{
				$\mtx{S}$ $\gets$ \StackMatrice($\mtx{S}$, $\mtx{S}^\prime$)
			}
	 	} 
		$\mtx{S} \gets$ \RandomSample($\mtx{S}$, $k$) \\
		\Return $\mtx{S}$
	}
\end{algorithm}

\subsection{SMOTE-PC}
\begin{figure}[!htbp]
	\centering
	\subfloat[Segment 0]{%
 		\includegraphics[width=0.2\textwidth]{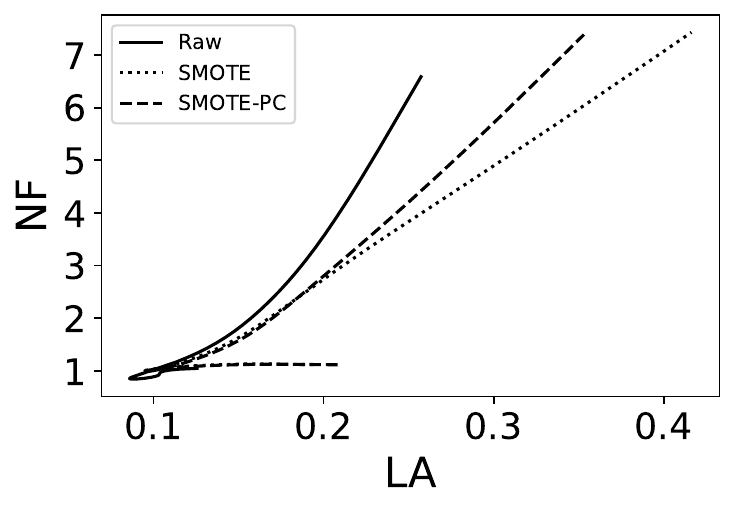}
		\label{fig::smotepc:distpcrq3:seg0}%
		}
	\hfil
	\subfloat[Segment 1]{%
 		\includegraphics[width=0.2\textwidth]{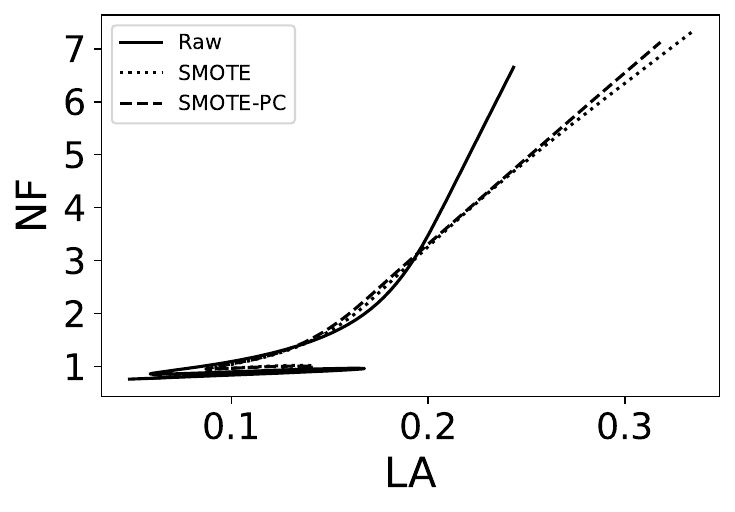}
		\label{fig::smotepc:distpcrq3:seg1}%
		}

	\subfloat[Segment 3]{%
 		\includegraphics[width=0.2\textwidth]{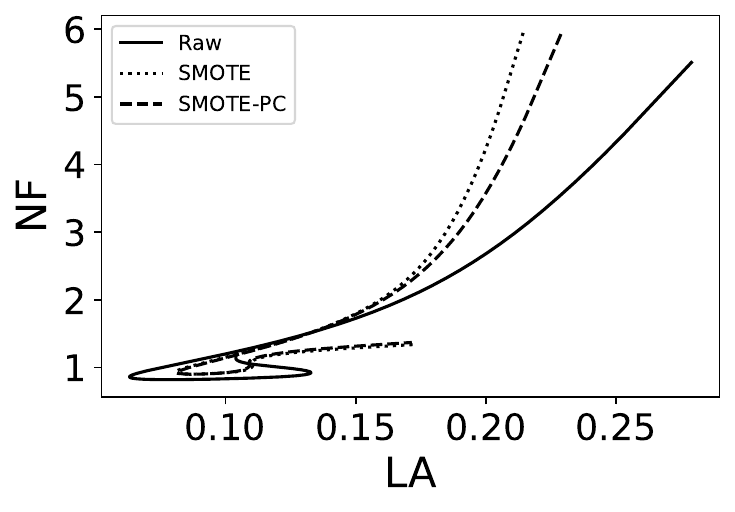}
		\label{fig::smotepc:distpcrq3:seg3}%
		}
	\hfil
	\subfloat[Segment 4]{%
 		\includegraphics[width=0.2\textwidth]{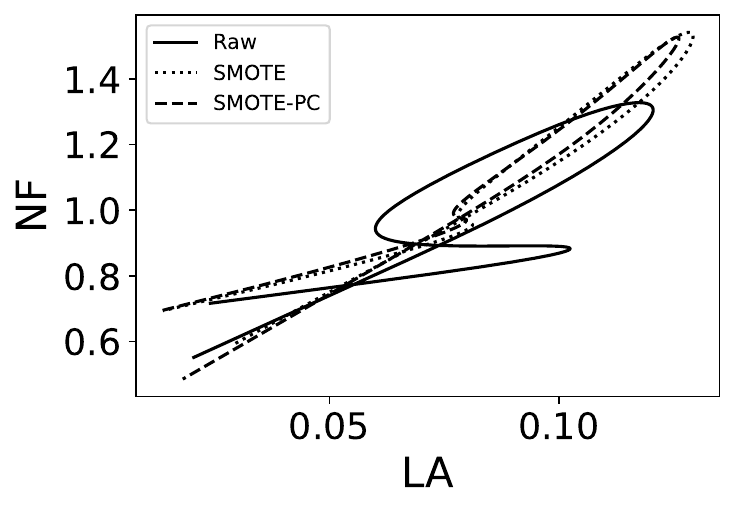}
		\label{fig::smotepc:distpcrq3:seg4}%
		}
	\caption{Principal curves of raw data, SMOTE balanced data and SMOTE-PC
	balanced data of 4 segments of the Eclipse Platform project}

	\label{fig:smotepc:distpcrq3}
\end{figure}

We propose a new class imbalance process technique called SMOTE-PC, i.e., SMOTE
enhanced with Principal Curves.  As shown in Algorithm~\ref{alg:smotepc}, the
idea of SMOTE-PC is as follows.  SMOTE randomly interpolate the minority class
data set to generate synthetic data instances for the minority class. Since
principal curves are non-linear summaries of the underlying distribution of the
data, we monitor whether adding SMOTE generated data instances to the raw
data set would alter the principal curves. We reject the generated data instances
when they cause a significant change in the principal curves.  In
Figure~\ref{fig:smotepc:distpcrq3}, we also compute and show the principal
curves of the segments of data set after we apply SMOTE-PC. The figure confirms
that the changes of principal curves introduced by SMOTE-PC are smaller than
those by SMOTE.

\section{Concept-Preserving Incremental Learning}
\label{sec:deepicp}
Based on the empirical results presented thus far and prior studies on
classification models for  JIT-SDP, we propose a concept-preserving incremental
learning framework for JIT-SDP and present a model of simple realization of the
framework called the Deep Incremental Concept-Presentation model (DeepICP). We
describe the framework and the DeepICP model in
Section~\ref{sec:deepicp:framework} and evaluate the model to answer RQ.4 in
Section~\ref{sec:deepicp:eval}.

\subsection{Design of Concept-Preserving Learning Framework}
\label{sec:deepicp:framework}
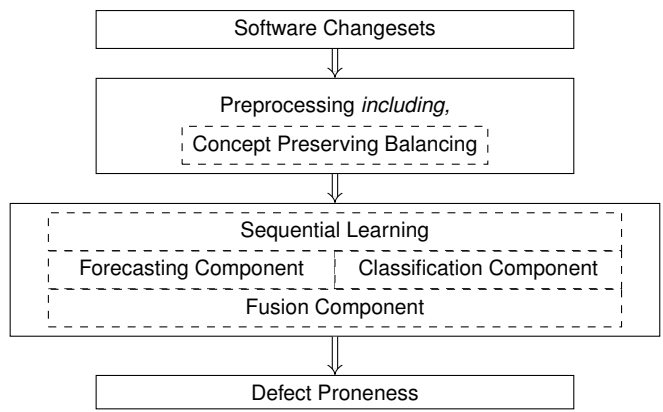
\begin{figure}[!htbp]
	\centering
	\begin{adjustbox}{max width=\columnwidth}
	\begin{tikzpicture}
		\tikzset{
			node distance=0.2in and 0.15in,
			every node/.style={
					text=black,
					fill=white,
					minimum width=2.5in,
					font=\footnotesize\sffamily
			},
			every path/.style={
					color=black,
			},
			align=center
		}

		\node(in)[rectangle, solid, draw]{Software Changesets};

		\node(prep)[rectangle, minimum width=1.6in,
			below=of in]{Preprocessing \textit{including,}};
		\node(cpb)[rectangle, 
			dashed,
			draw,
			minimum width=1.6in,
			yshift=0.2in,
			below=of prep]{Concept Preserving Balancing};
		\node(prepwrap)[rectangle, 
			solid, 
			draw, 
			draw opacity=1, 
			fill opacity=0,
			fit=(prep) (cpb)]{};

		\node(seq)[rectangle,
			dashed,
			draw,
			below=of prepwrap,
			minimum width=3in
			]{Sequential Learning};

		\node(auto)[rectangle, 
			dashed, 
			draw, 
			minimum width=1.5in,
			below of=seq,
			xshift=-0.75in
			]{Forecasting Component};
		\node(cls)[rectangle, 
			dashed, 
			draw, 
			minimum width=1.5in,
			below of=seq,
			xshift=0.75in
			]{Classification Component};
		\node(fusion)[rectangle, 
			dashed, 
			draw, 
			minimum width=3.0in,
			xshift=-0.75in,
			below of=cls]{Fusion Component};

		\node(modelwrap)[rectangle,
			solid,
			draw,
			draw opacity=1,
			fill opacity=0,
			minimum width=3.4in,
			fit=(seq) (auto) (cls) (fusion)]{};

		\node(out)[rectangle, solid, draw, 
			below=of modelwrap]{Defect Proneness};

		\draw[-{Implies},double equal sign distance, double](in) -- (prepwrap);
		\draw[-{Implies},double equal sign distance, double](prepwrap) -- (modelwrap);
		\draw[-{Implies},double equal sign distance, double](modelwrap) -- (out);

	\end{tikzpicture}
	\end{adjustbox}
	\caption{CPI-JIT framework of Concept-Preserving Sequential JIT-SDP Models}
	\label{fig:model:jitframework}
\end{figure}

We propose a Concept-Preserving Incremental learning framework for JIT-SDP
(CPI-JIT) that takes advantage of our understandings about
\begin{inparaenum}\item class imbalance processing, \item interdependent 
		relationship of
	software changesets; and \item prior advances in classification JIT-SDP
models.  \end{inparaenum} Figure~\ref{fig:model:jitframework} illustrates the
major components of the framework. 
		
We show empirically that class imbalance processing can have an impact on the
``concept'', or the underlying distribution of software changeset data and
since the regularities (or patterns) in the distribution are where a JIT-SDP
model learns to predict defective changes, we hypothesize that the class
imbalance processing can negatively impact learning of JIT-SDP models.
Following this, we propose to include a concept-preserving class imbalance
processing component in the framework.

We also observe that there are temporal interdependent relationships among
software changesets. Complement to prior classification-based JIT-SDP models,
we supplement the framework with a ``forecast'' component that extracts
information from preceding changesets of a target changeset by predicting the
characteristics of future changesets. Combining the outputs of the forecasting
component and the classification component, we predict the defectiveness status
of the target software changeset.

Prior JIT-SDP studies indicate that concept drift can negatively impact aged
JIT-SDP models. To address the problem, the studies propose three approaches,
\begin{inparaenum} \item by updating models with new chunks of software
	changesets data~\cite{tan2015online}, and \item by selecting recent training
data~\cite{mcintosh2017fix, bennin2020revisiting}, where chunks
are non-overlapping segments of JIT-SDP changesets; and by using an online
model~\cite{cabral2019class, tabassum2020investigation}.  \end{inparaenum} The
proposed framework is designed to encompass these approaches.  Incremental and
online models differs from offline models in that the incremental or online
models learn from the new incoming training examples by updating the models. If
we choose to update model by a large chunk, we can take advantage of prior
JIT-SDP models by retrain the model with the new chunk of the changeset data.
Although there is no consistent definition, literature generally refers to the
models that learn and update by chunks of new training examples as incremental
models, and those by one training example as online models.  If we elect to
update the model with a chunk of size 1, the incremental learning model becomes
an online model.

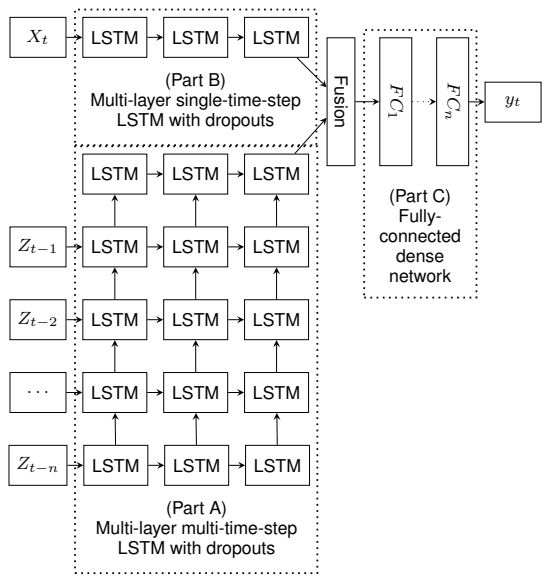
\begin{figure}[!htbp]
	\centering

	\begin{adjustbox}{max width=0.8\columnwidth}
	\begin{tikzpicture}
		\tikzset{
			node distance=0.20in and 0.10in,
			every node/.style={
					text=black,
					fill=white,
					font=\footnotesize\sffamily
			},
			every path/.style={
					color=black,
			},
			align=center,
			io/.style={
				rectangle,
				draw,
				minimum width=0.35in,
				minimum height=0.25in
			},
			lstm/.style={
				rectangle,
				draw,
				minimum width=0.40in,
				minimum height=0.25in
			},
			flow/.style={
				-stealth
			},
			flowdotted/.style={
				dotted,-stealth
			},
			note/.style={
				rectangle,
				align=center
			},
			wrapper/.style={
				rectangle,
				dotted,
				thick,
				draw,
				fill opacity=0,
				draw opacity=1,
				align=center
			},
			fusion/.style={
				rectangle,
				solid,
				draw,
				minimum width=0.80in,
				align=center,
				rotate=-90,
			},
			fc/.style={
				rectangle,
				solid,
				draw,
				minimum width=0.80in,
				minimum height=0.20in,
				rotate=-90,
				align=center
			}
		}

		\node(ft)[io]{$X_t$};
		\node(lstmt1)[lstm, right=of ft]{LSTM};
		\foreach \i/\j in {1/2, 2/3}
		{
			\node(lstmt\j)[lstm, right=of lstmt\i]{LSTM};
			\draw[flow](lstmt\i) -- (lstmt\j);
		}
		\draw[flow](ft) -- (lstmt1);
		\node(clstm)[note,below=of lstmt2, yshift=0.15in]{%
			(Part B)\\Multi-layer single-time-step \\LSTM with dropouts};
		\node[wrapper, fit=(lstmt1) (lstmt3) (clstm)]{};

		\node(emptyt0)[io, draw opacity=0, below=of ft, yshift=-0.40in]{};

		\node(ft1)[io, below=of emptyt0]{$Z_{t-1}$};
		\node(ft2)[io, below=of ft1]{$Z_{t-2}$};
		\node(fto)[io, below=of ft2]{$\ldots$};
		\node(ftn)[io, below=of fto]{$Z_{t-n}$};

		\node(lstmt10)[lstm, right=of emptyt0]{LSTM};
		\foreach \i in {1, 2, o, n}
		{
			\node(lstmt1\i)[lstm, right=of ft\i]{LSTM};
		}

		\foreach \i/\j in {n/o, o/2, 2/1, 1/0} {
			\draw[flow](lstmt1\i) -- (lstmt1\j);
		}

		\foreach \i in {1, 2, o, n}
		{
			\draw[flow](ft\i) -- (lstmt1\i);
		}

		\foreach \l/\i in {1/2, 2/3}
		{
			\foreach \j in {0, 1, 2, o, n}
			{
				\node(lstmt\i\j)[lstm, right=of lstmt\l\j]{LSTM};
			}
			\foreach \m/\n in {n/o, o/2, 2/1, 1/0} 
			{
				\draw[flow](lstmt\i\m) -- (lstmt\i\n);
			}
			\foreach \m in {n, o, 2, 1, 0} 
			{
				\draw[flow](lstmt\l\m) -- (lstmt\i\m);
			}
		}

		\node(seqlstm)[note,below=of lstmt2n, yshift=0.15in]{%
			(Part A)\\Multi-layer multi-time-step \\LSTM with dropouts};

		\node[wrapper, fit=(lstmt10) (lstmt3n) (seqlstm)]{};

		\node(fusion)[fusion, right=of lstmt3, yshift=0.10in]{Fusion};
		\draw[flow](lstmt30) -- (fusion);
		\draw[flow](lstmt3) -- (fusion);

		\node(fc1)[fc, above=of fusion, xshift=0.20in, yshift=0.15in]{$FC_1$};
		\node(fc2)[fc, above=of fc1, xshift=0.20in, yshift=0.15in]{$FC_n$};

		\node(classifier)[note,below=of fc2, xshift=-0.10in, yshift=-0.30in]{%
			(Part C)\\Fully-\\connected \\dense\\network};
		\node[wrapper, fit=(fc1) (fc2) (classifier)]{};

		\draw[flow](fusion) -- (fc1);
		\draw[flowdotted](fc1) -- (fc2);
		
		\node(out)[io, right=of fc2, yshift=0.40in, xshift=0.10in]{$y_t$};
		\draw[flow](fc2) -- (out);

	\end{tikzpicture}
	\end{adjustbox}
	\caption{A Concept-Preserving Incremental JIT-SDP Model built from 
	Deep Neural Networks (DeepICP). $X_t$ is the feature vector
	of the software changeset at time t and its defectiveness status $y_t$
	is to be determined. $Z_{t-1}$, $\ldots$, $Z_{t-n}$ are the feature
	vectors and their known defectiveness status of the $n$ preceding
	software changesets, i.e., $Z_i = (X_i, y_i)$.
	}
	\label{fig:model:deepicp}
\end{figure}

As a proof of concept, Figure~\ref{fig:model:deepicp} presents a simple
realization of this framework.  We call the JIT-SDP model presented in the
figure the Deep Incremental Concept-Preserving model (DeepICP) that takes
advantage of the interdependent relationship of software changesets by
integrating an autoregressive model shown as Part A in
Figure~\ref{fig:model:deepicp}. We design this part of the model using a Long
Short-Term Memory (LSTM) neural network as in Section~\ref{sec:relation:auto}.
Leveraging prior JIT-SDP models, we adopt the design of the JIT-SDP
classification model, as shown in Part B of the model.  The outputs of both
Parts A and B are combined and given to a fully-connected dense neural network
that acts as a classifier (Part C). To avoid computational expensive hyperparameter
tuning including network structural parameters, we borrow parameters from prior
research~\cite{abuhamad2018large}.

\subsection{Evaluation}
\label{sec:deepicp:eval}
In order to evaluate the proposed model and answer RQ.4, we break down RQ.4
into 5 subquestions.
\begin{enumerate}
	\item (RQ.4a) Does SMOTE-PC help the model achieve better performance on
		predictive defective changes? 

	\item (RQ.4b) Does the forecasting component of the model help the model
		achieve better performance on predictive defective changes? 

	\item (RQ.4c) Does the ``age'' of the DeepICP model impact its performance on
		predict defective changes? We define the model age as the arrival time
		between the oldest changeset in the test data set and the newest changeset
		in the training data set. 

	\item (RQ.4d) Does incremental learning or model update over time improve the
		model's predictive performance on defective changes? 

	\item (RQ.4e) How well does DeepICP predict defective changes when compared
		to the logistic regression model proposed by Kamei et
		al.~\shortcite{kamei2012large} that we select as a baseline model?
\end{enumerate}

The proposed CPI-JIT framework begins with a concept-preserving balancing
processing.  To show the benefit of this component, RQ.4a is to test on whether
there is a substantial benefit from using SMOTE-PC, a concept-preserving
balancing processing technique by using the proposed DeepICP model as an
instrument. Next, realizing that although the changesets preceding to the
target set contains useful information for defect prediction, the information
may also be noisy, we are to answer whether the DeepICP model can benefit from
the useful but noisy information in the preceding changesets in RQ.4b.  The
intention of RQ.4c is to confirm prior studies about the impact of concept
drift in software changeset data on JIT-SDP models.  RQ.4d is to demonstrate
how the proposed model can adapt to concept drift.  RQ.4e is to show how well
the DeepICP model, a simple realization of the CPI-JIT framework performance
when compared to the model in Kamei et al.~\shortcite{kamei2012large}.  We
design experiments for each of these questions and describe the experiment
settings as we answer these questions in Sections
from~\ref{sec:deepicp:ablation:smotepc} to~\ref{sec:deepicp:baseline}.

Throughout this section, we evaluate two performance metrics, AUC-ROC and F1.
AUC-ROC, the Area Under the Curve of Receiver Operating Characteristic is a
threshold-free performance metric and F1 is the harmonic mean of Precision and
Recall, i.e., $F1 = 2 \cdot \text{Precision} \cdot
\text{Recall}/(\text{Precision} + \text{Recall})$. AUC-ROC and F1 are the most
widely used evaluation metrics in software defection prediction
studies~\cite{hall2011systematic, zhao2023systematic}. These two offer an
advantage when compared to the others. Because AUC is threshold-free, it can not
be manipulated by changing decision threshold. F1 is a harmonic mean of recall
and precision, although threshold-dependent, it is difficult to manipulate
because when one of recall and precision often improves at the expense of the
other. 

\begin{table}[!htbp]
	\centering
	\caption{AUC and F1 of DeepICP with SMOTE and with SMOTE-PC}
	\label{tab:model:smotepc:t0}
	\begin{adjustbox}{max width=1\columnwidth}
	\begin{tabular}{l r r r r r r }
		\toprule
						& \multicolumn{3}{c}{AUC} & \multicolumn{3}{c}{F1} \\
    Project & SM & SM-PC &  I\%   & SM & SM-PC & I\% \\
    \midrule
    JDT  & 0.69  & 0.73  &  5.8\% & 0.32 & 0.46 &  43.8\% \\
    PLA  & 0.67  & 0.75  & 11.2\% & 0.37 & 0.42 &  13.5\% \\
    MOZ  & 0.64  & 0.81  & 26.6\% & 0.13 & 0.36 & 176.9\% \\
    POS  & 0.76  & 0.80  &  5.3\% & 0.56 & 0.58 &   3.6\% \\
    \midrule
    AMQ  & 0.78  & 0.78 &   0.0\% & 0.66 & 0.66 &   0.0\% \\
    CAM  & 0.83  & 0.83 &   0.0\% & 0.14 & 0.35 & 150.0\% \\
    HDP  & 0.64  & 0.72 &  12.5\% & 0.10 & 0.20 & 100.0\% \\
    HBA  & 0.68  & 0.71 &   4.4\% & 0.38 & 0.46 &  21.1\% \\
    MAH  & 0.81  & 0.85 &   4.9\% & 0.60 & 0.66 &  10.0\% \\
    JPA  & 0.69  & 0.78 &  13.0\% & 0.34 & 0.39 &  14.7\% \\
		\bottomrule
 		\multicolumn{7}{p{3.0in}}{%
 			\footnotesize{We use acronyms SM and SM-PC for SMOTE and SMOTE-PC, 
			and I\% for Improvement\%.}
 		}
	\end{tabular}
	\end{adjustbox}
\end{table}

\begin{figure}[!htbp]
	\centering
	\subfloat[AUC-ROC]{
		\includegraphics[width=0.45\columnwidth]{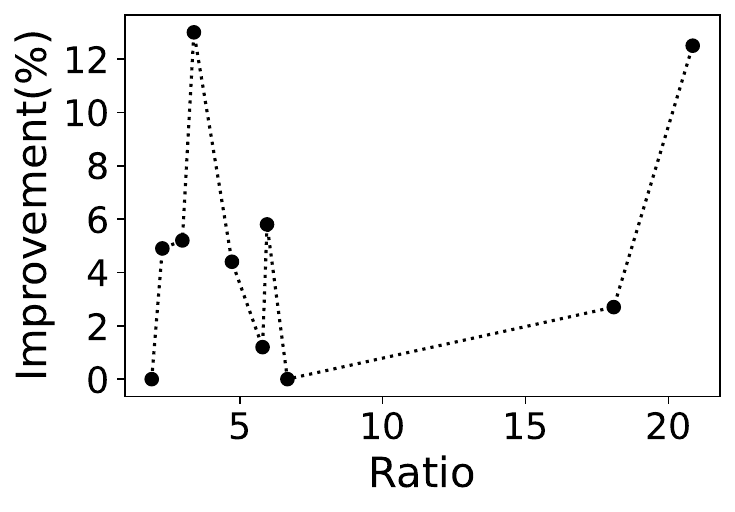}
		}
	\hfil
	\subfloat[F1]{
		\includegraphics[width=0.45\columnwidth]{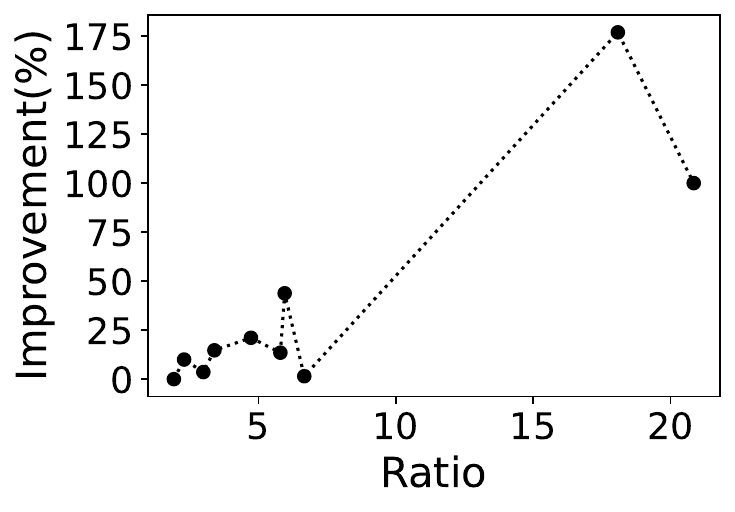}
		}
	\caption{Improvements of AUC-ROC and F1 of the model with SMOTE-PC over
	those with SMOTE versus clean-to-defect changeset ratios}
	\label{fig:model:smotepc:t0:ratio}
\end{figure}

\subsection{Effects of SMOTE-PC}
\label{sec:deepicp:ablation:smotepc}

\subsubsection{Experiment Setting}
To answer RQ.4a, we order and then divide the changesets of each project into
20 segments numbered as segment 0 to segment 19.  We structure this experiment
as an ablation study where we examine the model with and without SMOTE-PC. When
without SMOTE-PC, we use ordinary SMOTE.  We use 4 consecutive segments of
changeset data beginning at segment 8 as the training and validation data,
among which, first 90\% of the changesets are the training data instances and
10\% are for validation.  The test data set is the segment after the training
and validation data.  We evaluate the trained model using this test data set.
For convenience, we denote the training and validation data set as $D_t$, and
the test data sets as segment $D_{t+0}$,

\begin{table}[!htbp]
	\centering
	\caption{Correlations of Improvement of AUC-ROC and F1 versus Clean-to-Defect Changeset Ratio.}
	\label{tab:model:smotepc:t0:ratio}
	\begin{tabular}{r r r r}
	\toprule
		\multicolumn{2}{c}{Improvement of AUC-ROC vs. Ratio} & \multicolumn{2}{c}{Improvement of F1 vs. Ratio} \\
		Coefficient & p-value  & Coefficient & p-value \\
	\midrule
		0.30 & 0.41 & 0.88 & 7.2E-4 \\
	\bottomrule
	\end{tabular}
\end{table}

\subsubsection{Results}

Table~\ref{tab:model:smotepc:t0} is the evaluation results on test segment
$D_{t+0}$. The table shows that there is a cross-board improvement
of AUC-ROC and F1 scores of the model with SMOTE-PC over those of the model
with SMOTE. The results in the table
also show that the improvement of F1 has much higher magnitude than that of
AUC-ROC. This is often a result that ROC curves intersect, which is not
uncommon in classification studies~\cite{pencina2008evaluating}.

SMOTE and SMOTE-PC generates synthetic training instances of minority class instances to
balance the training data set. We expect SMOTE-PC introduces less severe
changes in the underlying distribution of data that SMOTE does. In particular,
SMOTE and SMOTE-PC would generate more synthetic changeset instances of the
minority class, i.e., the defect-inducing changesets when the clean-to-defect
changeset ratio increases. As the amount of the synthetic
changeset instances increases, we expect SMOTE-PC fares well than SMOTE
in preserving the underlying data distribution. To test on this
understanding, we investigate
whether there is a correlation between the improvements of AUC-ROC and F1
scores of the model with SMOTE-PC over those with SMOTE and the defect-to-clean
changeset ratios.

Figure~\ref{fig:model:smotepc:t0:ratio} plots the improvement ratios between
SMOTE-PC and SMOTE of the \numprojs software projects with varying
clean-to-defect changeset ratios. Table~\ref{tab:model:smotepc:t0:ratio} shows
their correlation coefficients.
The figure and the table show the followings.
There is no apparent relationship between the improvements of AUC-ROC
and the projects' clean-to-defect
changeset ratios.  In contrast, the correlation between the improvements of F1 scores
and the clean-to-defect changeset ratios is strong. Our interpretation to
the observation is that SMOTE-PC can preserve the data distribution of the
minority class, i.e., the defect-inducing changesets, thereby increase the
model's ability to predict on the minority class, and leads to the improvement
of F1 scores; however, it is less so for AUC-ROC as test data sets are dominated
by clean changesets.

\begin{table}[!htbp]
	\centering
	\caption{AUC and F1 of DeepICP {\em with} Forecasting and {\em without} Forecasting}
	\label{fig:model:forecast:t0}
	\begin{adjustbox}{max width=1\columnwidth}
 	\begin{tabular}{p{0.30in}%
 		>{\raggedleft\arraybackslash}p{0.31in} >{\raggedleft\arraybackslash}p{0.29in} >{\raggedleft\arraybackslash}p{0.28in}%
 		>{\raggedleft\arraybackslash}p{0.31in} >{\raggedleft\arraybackslash}p{0.29in} >{\raggedleft\arraybackslash}p{0.33in}}
		\toprule
						& \multicolumn{3}{c}{AUC} & \multicolumn{3}{c}{F1} \\
		Project & No-FC & FC &  I\% & No-FC & FC & I\% \\
		\midrule
		JDT  & 0.73  &0.73   & 0.0\%  &0.46  & 0.44 & -4.3\% \\
		PLA  & 0.75  &0.77   & 2.7\%  &0.42  & 0.49 & 16.7\% \\
		MOZ  & 0.81  &0.83   & 2.5\%  &0.36  & 0.36 &  0.0\% \\
		POS  & 0.79  &0.79   & 0.0\%  &0.58  & 0.58 &  0.0\% \\
		\midrule
		AMQ  &0.79   &0.78   &-1.3\%  &0.69  &0.67 &  -2.9\% \\
		CAM  &0.83   &0.83   & 0.0\%  &0.27  &0.32 &  18.5\% \\
		HDP  &0.72   & 0.77  & 6.9\%  &0.20  &0.18 &  -1.0\% \\
		HBA  &0.71   & 0.70  & 1.4\%  &0.46  &0.45 &  -2.2\% \\
		MAH  &0.85   &0.85   & 0.0\%  &0.66  &0.66 &   0.0\% \\
		JPA  &0.77   &0.77   & 0.0\%  &0.39  &0.40 &   2.6\% \\
		\bottomrule
 		\multicolumn{7}{p{3.2in}}{%
 			\footnotesize{We use acronyms No-FC and FC for the case without
			the forecasting component and that with forecasting component in 
			the model. Like before, we use I\% for Improvement\%.}
 		}
	\end{tabular}
	\end{adjustbox}

\end{table}

\subsection{Effects of Forecasting Component}
\label{sec:deepicp:ablation:forecast}

\subsubsection{Experiment Setting}
This is to answer RQ.4b.  Similar to
Section~\ref{sec:deepicp:ablation:smotepc}, this is also an ablation study
where we turn on and off the forecasting component in the model while keeping
SMOTE-PC.  The rest of the setting is identical to
Section~\ref{sec:deepicp:ablation:smotepc}.

\subsubsection{Results}
Table~\ref{fig:model:forecast:t0} show the AUC-ROC and F1 scores of the model
with the forecasting component and those of without.  Among 1 of \numprojs
projects, we see a decrease of AUC-ROC, among 5 of \numprojs no effects, but 4
of \numprojs an improvement. Among 3 of \numprojs projects, there is an
improvement of F1 scores, 3 of \numprojs no effect, and 4 of \numprojs a
decrease. The results are thus mixed, and for some projects forecasting
component helps while for the others it hurts. The mixed results are likely the
result that the preceding changesets to the target changeset although contain
useful information, but also noise.

\begin{table}[!htbp]
	\centering
	\caption{AUC-ROC and F1 of DeepICP {\em without} model update for 4 consecutive
	test segments}
	\label{tab:model:cmp:t0_3}
	\begin{adjustbox}{max width=1\columnwidth}
	\begin{tabular}{l r r r r | r r r r}
		\toprule
		        & \multicolumn{4}{c}{AUC} & \multicolumn{4}{c}{F1} \\
		Project & \footnotesize{$D_{t+0}$} & \footnotesize{$D_{t+1}$} & \footnotesize{$D_{t+2}$} & \footnotesize{$D_{t+3}$}
						& \footnotesize{$D_{t+0}$} & \footnotesize{$D_{t+1}$} & \footnotesize{$D_{t+2}$} & \footnotesize{$D_{t+3}$}\\
						
		\midrule
    JDT  & 0.73  & 0.73  & 0.73 & 0.80 & 0.46 & 0.43 & 0.42 & 0.41 \\
    PLA  & 0.75  & 0.71  & 0.78 & 0.75 & 0.42 & 0.38 & 0.38 & 0.42 \\
    MOZ  & 0.81  & 0.82  & 0.79 & 0.82 & 0.36 & 0.32 & 0.35 & 0.30 \\
    POS  & 0.79  & 0.85  & 0.58 & 0.87 & 0.58 & 0.59 & 0.60 & 0.62 \\
    \midrule
    AMQ  & 0.79  & 0.66  & 0.77 & 0.76 & 0.69 & 0.43 & 0.60 & 0.57 \\ 
    CAM  & 0.83  & 0.81  & 0.79 & 0.81 & 0.35 & 0.25 & 0.27 & 0.27 \\
    HDP  & 0.72  & 0.62  & 0.65 & 0.65 & 0.20 & 0.16 & 0.13 & 0.08 \\
    HBA  & 0.71  & 0.74  & 0.72 & 0.74 & 0.46 & 0.51 & 0.51 & 0.49 \\
    MAH  & 0.85  & 0.82  & 0.78 & 0.78 & 0.66 & 0.76 & 0.48 & 0.56 \\
    JPA  & 0.78  & 0.75  & 0.69 & 0.59 & 0.39 & 0.33 & 0.30 & 0.20 \\
		\bottomrule
 		\multicolumn{9}{p{3.8in}}{%
			\scriptsize{ We denote the 4 consecutive segments of test data sets as
			segment $D_{t+0}$, $D_{t+1}$, $D_{t+2}$, and $D_{t+3}$, and use ``AVG''
			as  acronyms for ``Average'.}
 		}
	\end{tabular}
	\end{adjustbox}

\end{table}

\subsection{Concept Drift and Age of Model}
\label{sec:model:age}

\subsubsection{Experiment Setting}
This is to answer RQ.4c.  The experiment setting is identical to
Section~\ref{sec:deepicp:ablation:smotepc} when it comes to the training of the
model except only with SMOTE-PC.  We evaluate the trained model using segments
after the training and validation data set.  Similar to before, we denote the
training data set as $D_t$, and the test data sets as segment $D_{t+0}$,
$D_{t+1}$, $D_{t+2}$, and $D_{t+3}$ where $D_{t+0}$ is the segment immediately
follows $D_t$ and $D_{t+(i+1)}$ is the next segment after $D_{t+i}$, $i=0, 1,
2$.  For this experiment, we do not update the model as we move from $D_{t+0}$
to $D_{t+3}$. In this setting, the trained model ages as we move from $D_{t+0}$
to $D_{t+3}$.

\subsubsection{Results}

Table~\ref{tab:model:cmp:t0_3} is the evaluation results. When examining the
average AUC-ROC and F1 scores, we observe that there is indeed a degraded
predictive performance as the trained model ages, although it is more
consistent in terms of F1 scores than AUC-ROC. However, we cannot observe
this consistently when each individual software project is concerned. From the
table, we see that among 7 of \numprojs software projects, the F1 scores of test
segments $D_{t+i}$, $i=1,2,3$ are consistently lower than that of $D_{t+0}$.
These projects are JDT, PLA, MOZ, AMQ, CAM, HDP, and JPA. When we examine
AUC-ROC, we only have this observation among 4 software projects of \numprojs
projects.  These projects are CAM, HDP, MAH, and JPA.

\begin{table}[!htbp]
	\centering
	\caption{AUC-ROC of DeepICP {\em without} and {\em with} model update for 4 consecutive
	test segments}
	\label{tab:model:update:t0_3:auc}
	\begin{adjustbox}{max width=1\columnwidth}
	\begin{tabular}{l r r c r r c r r c}
		\toprule
		Project 
			& $D_{t+1}$ & $D_{t+1}^\prime$  
			& $D_{t+2}$ & $D_{t+2}^\prime$  
			& $D_{t+3}$ & $D_{t+3}^\prime$  \\
		\midrule
		JDT  & 0.73  &0.75 \mplus  & 0.73 & 0.74 \mplus & 0.80 & 0.80 \meq   \\
		PLA  & 0.71  &0.72 \mplus  & 0.78 & 0.78 \meq   & 0.75 & 0.76 \mplus \\
		MOZ  & 0.85  &0.82 \mminus & 0.82 & 0.80 \mminus& 0.83 & 0.79 \mminus\\
		POS  & 0.85  &0.84 \mminus & 0.83 & 0.82 \mminus& 0.87 & 0.88 \mplus \\
		\midrule
		AMQ  & 0.66  &0.70 \mplus  & 0.77 & 0.79 \mplus & 0.76 & 0.79 \mplus \\ 
		CAM  & 0.81  &0.86 \mplus  & 0.79 & 0.82 \mplus & 0.81 & 0.84 \mplus \\
		HDP  & 0.62  &0.68 \mplus  & 0.65 & 0.70 \mplus & 0.65 & 0.74 \mplus \\
		HBA  & 0.74  &0.75 \mplus  & 0.72 & 0.73 \mplus & 0.74 & 0.75 \mplus \\
		MAH  & 0.82  &0.82 \meq    & 0.78 & 0.74 \mminus& 0.78 & 0.72 \mminus\\
		JPA  & 0.75  &0.77 \mplus  & 0.69 & 0.71 \mplus & 0.59 & 0.62 \mplus \\
		\bottomrule
 		\multicolumn{7}{p{3.2in}}{%
			\scriptsize{
				We
				use $D_{t+i}$ to indicate AUC-ROC of the
				model {\em without} model update tested on segment $D_{t+i}$ while
				$D_{t+i}^\prime$ {\em with} model update tested on $D_{t+i}$, where
				$i=1, 2, 3$. We use ``(+)'' , ``(-)'', and ``(=)'' in the table to
				indicate whether the model update has a positive impact, a negative
				impact, or no impact on the predictive performance of the model.
			} 
		}
	\end{tabular}
	\end{adjustbox}
	\vskip -1em
\end{table}

\subsection{Effect of Incremental Learning}
\label{sec:model:incremental}

\subsubsection{Experiment Setting}
This is to answer RQ.4d. The experimental setting is similar to
Section~\ref{sec:model:age}, but with a difference, i.e.,
after the training and validation using data set $D_t$, we update the model by
retraining the model using the newly arrived segment of changesets, i.e.,
segment $D_{t+0}$. We then compare the predictive performance of the model with
and without the model update using consecutive segments of test data after the
$D_{t+0}$, i.e., segments $D_{t+i}$ where $i=1, 2, 3$. 

\subsubsection{Results}

Table~\ref{tab:model:update:t0_3:auc} compares AUC-ROC of the models in two
scenarios, with and without model update while
Table~\ref{fig:model:update:t0_3:f1} contrasts F1 scores.

Examining Table~\ref{tab:model:update:t0_3:auc}, we observe among \numprojs software
projects $\times$ 3 $ = 30$ pairs of comparisons of AUC-ROC scores with and
without model update, 19 or 63\% pairs have improved AUC-ROC. When it comes to
F1 scores, 16 or 53\% pairs see an improvement of F1 scores. Although there is
no consistent improvement of predictive performance when we update the models,
our results are consistent with prior studies. For instance, Tan et
al.\shortcite{tan2015online} found that model updates only improve the
predictive performance of a model under certain circumstances.

\begin{table}[!htbp]
	\centering
	\caption{F1 scores of DeepICP {\em without} and {\em with} model update for 4 consecutive
	test segments}
	\label{fig:model:update:t0_3:f1}
	\begin{adjustbox}{max width=1\columnwidth}
	\begin{tabular}{l r r r r r r r}
		\toprule
		Project 
			& $D_{t+1}$ & $D_{t+1}^\prime$ 
			& $D_{t+2}$ & $D_{t+2}^\prime$ 
			& $D_{t+3}$ & $D_{t+3}^\prime$  \\
		\midrule
		JDT  & 0.43  & 0.44 \mplus  & 0.42 & 0.45 \mplus  & 0.41 & 0.43 \mplus \\
		PLA  & 0.38  & 0.44 \mplus  & 0.38 & 0.44 \mplus  & 0.42 & 0.50 \mplus \\
		MOZ  & 0.34  & 0.31 \mminus & 0.37 & 0.32 \mminus & 0.32 & 0.25 \mminus\\
		POS  & 0.59  & 0.54 \mminus & 0.60 & 0.57 \mminus & 0.62 & 0.60 \mminus\\
		\midrule
		AMQ  & 0.43  & 0.46 \mplus  & 0.60 & 0.59 \mminus & 0.57 & 0.57 \meq   \\ 
		CAM  & 0.25  & 0.33 \mplus  & 0.27 & 0.31 \mplus  & 0.27 & 0.32 \mplus \\
		HDP  & 0.16  & 0.20 \mplus  & 0.13 & 0.16 \mplus  & 0.08 & 0.15 \mplus \\
		HBA  & 0.51  & 0.51 \meq    & 0.51 & 0.50 \mminus & 0.49 & 0.49 \meq   \\
		MAH  & 0.76  & 0.73 \mminus & 0.48 & 0.44 \mminus & 0.56 & 0.51 \mminus\\
		JPA  & 0.33  & 0.37 \mplus  & 0.30 & 0.33 \mplus  & 0.20 & 0.22 \mplus \\
		\bottomrule
 		\multicolumn{7}{p{3.2in}}{%
			\scriptsize{
				The notations in this table are identical to
				Table~\ref{tab:model:update:t0_3:auc} except that we show F1 scores
				here. 
			} 
 		}
	\end{tabular}
	\end{adjustbox}
\end{table}

\subsection{Baseline Model Comparison}
\label{sec:deepicp:baseline}

\subsubsection{Experiment Setting}
This is to answer RQ.4e.  Experiment setting is identical to
Section~\ref{sec:deepicp:ablation:smotepc} except that we train and evaluate the DeepICP
model and the baseline model with SMOTE-PC. The baseline model is the model in
Kamei et al.~\shortcite{kamei2012large} but with SMOTE-PC.

\begin{table}[!htbp]
	\centering
	\caption{AUC-ROC and F1 of DeepICP {\em without} model update and the baseline
	model.}
	\label{tab:model:cmp:t0}
 	\begin{tabular}{p{0.30in}%
 		>{\raggedleft\arraybackslash}p{0.29in} >{\raggedleft\arraybackslash}p{0.29in} >{\raggedleft\arraybackslash}p{0.28in}%
 		| >{\raggedleft\arraybackslash}p{0.29in} >{\raggedleft\arraybackslash}p{0.29in} >{\raggedleft\arraybackslash}p{0.33in}}
		\toprule
						& \multicolumn{3}{c}{AUC} 
						& \multicolumn{3}{c}{F1}\\
		Project & B-L & D-I & I\% & B-L & D-I & I\% \\
		\midrule
        JDT & 0.74   & 0.73   & -1.4\%   & 0.45  & 0.46 &   2.2\% \\
        PLA & 0.73   & 0.75   &  2.7\%   & 0.45  & 0.42 &  -6.7\% \\
        MOZ & 0.79   & 0.81   &  2.5\%   & 0.26  & 0.36 &   3.9\% \\
        POS & 0.76   & 0.80   &  5.3\%   & 0.56  & 0.58 &   3.6\% \\
        \midrule
        AMQ & 0.78   & 0.79   &  1.3\%   & 0.68  & 0.69 &   1.5\% \\ 
        CAM & 0.83   & 0.83   &  0.0\%   & 0.32  & 0.35 &   9.4\% \\
        HDP & 0.79   & 0.72   & -8.7\%   & 0.14  & 0.20 &  42.9\% \\
        HBA & 0.71   & 0.71   &  0.0\%   & 0.44  & 0.46 &   4.5\% \\
        MAH & 0.84   & 0.85   &  1.2\%   & 0.61  & 0.66 &   8.2\% \\
        JPA & 0.77   & 0.78   &  1.3\%   & 0.36  & 0.39 &   8.3\% \\
				\midrule
				AVG & 0.77 	 & 0.78	  & 0.4\%	   & 0.43	 & 0.46	&   9.1\%\\

		\bottomrule
		\multicolumn{7}{p{3.0in}}{%
			\footnotesize{
			We use ``AVG`` for Average, and ``B-L'', ``D-I'', and I\% as acronyms for ``Baseline'', ``DeepICP'',
			and Improvement\%.}
		}
	\end{tabular}

\end{table}

\subsubsection{Results}

Table~\ref{tab:model:cmp:t0} is the evaluation results. In terms of F1, DeepICP
outperforms the baseline model for 9 out of \numprojs software projects. In
terms of AUC-ROC, DeepICP outperforms the baseline model in 6 software
projects, i.e., projects PLA, MOZ, POS, AMQ, MAH, and JPA while it has equal or
inferior predictive performance in 4 projects, projects JDT, CAM, HDP, and HBA.

\section{Threats to Validity}
\label{sec:threats}
The categories of threats to validity of this study that we discuss below are
internal validity, construct validity, and external validity

\textbf{Internal Validity} This work rests on the analysis and the observation
on the interdependent relationship of software changesets and their
defectiveness status. To lessen the threats to validity, we assemble two data
sets, one is the data set shared by Kamei et al.\shortcite{kamei2012large} that
numerous studies have examined, and the other is the data set of
\numapacheprojs open-source projects collected by the authors of this article. 

Another concern is the defectiveness status of the dataset. To control the
quality, we select the open-source projects that have been consistently using
an ITS to manage issue reports and link issue reports in
their SCM commit messages~\cite{fan2019impact}.  Following this, we select an
SZZ implementation that has gone through a rigorous evaluation and use it to
identify defect-inducing changesets~\cite{rosa2021evaluating}.  This allows us
to maintain the quality of fixing changeset identification. 

Nevertheless, it is possible that there are misidentified defect-inducing
changesets. Future research should consider this limitation, such as, examine
how sensitive the results here are to the identification errors of
defect-inducing changesets. 

Additionally, there is another challenge of dealing with verification
latency~\cite{cabral2019class}. A typical approach to handle this is to set up a
fixed gap between training dataset and evaluation dataset, and the gap is not
actual latencies. This can have an impact on the validity of this work.

Aside in the above, training of deep neural networks can be stochastic due to
initial weights and training algorithms. Our results are from multiple segments
per project for 10 projects, as such, we consider this threat although exists,
not a significant one.

\textbf{External Validity}
Having considered threats that limit the ability to generalize the results of
this study, we select and experiment with \numprojs software projects that are of
different application domains, of different sizes, and span different
development periods as shown in Table~\ref{tab:datasets}.

Our conclusion is that the interdependent relationship and concept drift of
software changesets are important factors to consider when it comes to design
incremental learning models for JIT-SDP; however, the models respond
differently for different software projects and in different experiment
settings. We derive this conclusion from experimenting on the \numprojs software
projects. Nonetheless, additional replication studies are necessary to verify
the results.

\textbf{Construct Validity}
To reduce the threats to the design of our experiments, we select feature
representation of software changesets as the 14 software change metrics by
Kamei et al.~\shortcite{kamei2012large}, and these 14 software change metrics
are widely used in JIT-SDP studies, and thus can be considered as a
``standard''.  

To evaluate the prediction performance of the JIT-SDP models, we select AUC-ROC
and F1 score. AUC-ROC is a threshold-free evaluation criterion while F1-score is
a harmonic mean of Precision and Recall. Both of these are different to
manipulate, and are frequently used to evaluate JIT-SDP models.

To characterize the interdependent relationship of software changesets, we
use joint probability distribution and conditional probability distribution,
which are fundamental concepts in statistics. To quantify concept drift the
underlying  probability of the data, we select principal curve that has also
been examined in numerous studies in the past decade. 

\textbf{Replication Package}
Finally, we recognize that there still be unmitigated threats to validity.
Replications of this study using the same data sets and additional data sets
are necessary to ascertain whether our conclusions continue to hold, for other
software projects and in the future. To facilitate replication, we provide a
replication package that consists of the software code and the data set. The
replication package is at
\url{https://github.com/huichen-cs/deepicpreplication.git}.

\section{Conclusion and Future work}
\label{sec:sum}
In this work, we investigate the relationship of software changesets, and
characterize concept drift in software changesets, which leads to the proposal
of an incremental learning framework and a model for JIT-SDP. Via a case study
of \numprojs open-source projects, we reach the following observations.

\begin{enumerate}
	\item Software changesets are not independent events. There are temporal
		interdependent relationship among software changesets. The pattern in the
		relationship can aid defect prediction for JIT-SDP among software
		evaluated software projects. 

	\item Concept drift exists in software changesets. Past researches examine
		impacts of concept drifts on the predictive performance of JIT-SDP models,
		and also design online models to adapt to the class imbalance evolution, a
		source of concept drift.  We propose to use principal curves to
		characterize the source of concept drift. Using principal curves, we
		observe that software projects can exhibit concept drift not obvious from
		class imbalance evolution; and also observe software changesets can have a
		concept drift indicated by the class imbalance status, but unseen in
		principal curves.  We thus argue that changes of principal curves represent
		a new source of concept drift. 

	\item To illustrate the benefit of adapting a JIT-SDP model to the concept
		drift characterization via principal curve, we propose a class balancing
		technique called SMOTE-PC.  Using principal curves, SMOTE-PC attempts to
		preserve the distribution of the data while balancing the software
		changeset data. We hypothesize that the distribution of the changeset data
		contains the very regularities that JIT-SDP models learn to predict
		defective changes, and the change of the underlying distribution of the
		data can negatively impact the JIT-SDP models.  The experimental results
		show that SMOTE-PC introduces less changes on the distribution of data than
		SMOTE, and the JIT-SDP model using SMOTE-PC outperforms SMOTE in terms of
		F1 scores consistently. The results validate our method of characterizing
		the underlying distribution of data using Principal Curves.

	\item Finally, we consider the above findings and propose an incremental
		learning framework for JIT-SDP. In the framework, we propose the idea of
		concept-preserving preprocessing where we should aim to reduce changes in
		the ``concepts'', i.e., the underlying distribution of data. We also
		propose to take advantage of the temporal interdependent relationship of
		the software changes to aid defect prediction, i.e., we should also
		consider  forecasting models in addition to classification models.  We
		illustrate this framework by realizing the ideas in an incremental learning
		model called DeepICP. DeepICP uses SMOTE-PC to preserve the ``concepts''
		embedded in the changeset data.  It combines both a forecast component, an
		auto-regressive model and a classification component. The model adapts to
		concept drifts incrementally.

\end{enumerate}

These observations lead us to believe that concept-preserving and integrating
both forecasting and classification models can benefit JIT-SDP predictive performance
among some software projects. Researchers and practitioners may want to 
evaluate these in their models in order to produce a superior JIT-SDP system, 
in particular, in an incremental learning framework.

Finally, we would like to point out several limitations and future work
directions. First, DeepICP is a simple realization of the proposed framework and
clearly needs improvement in its design. %
Second, our model does not produce cross-board
improvement for all software projects evaluated. The question that under which
condition a software project can benefit from our JIT-SDP model is yet to be
answered. Third, our work only focuses on an oversampling approach that balances
training data. We are also interested in investigating the methodology presented in
this work, in particular, characterizing data distribution using Principle
Curves and combining forecasting and classification models in other two
categories of data balancing approaches, the undersampling and modeling
approaches~\cite{zhao2023systematic}.

\balance

\bibliographystyle{IEEEtran}
\bibliography{jitsdp,jitsdp_snowball,rlsesdp,ml,se,pc,szz,self,stats}

\end{document}